\documentclass[11pt,letterpaper]{article}

\usepackage[margin=3cm]{geometry}

\usepackage{amsfonts}
\usepackage{amssymb}
\usepackage{mathrsfs}
\usepackage{amsmath,amssymb}
\usepackage{bbm}

\usepackage{tabularx}
\newcommand{\PreserveBackslash}[1]{\let\temp=\\#1\let\\=\tem}

\makeatletter\renewcommand{\section}{\@startsection
{section}{1}{\z@}{-2.5ex plus -1ex minus
    -.2ex}{2.3ex plus .2ex}{\centering\large\sc }}

\makeatletter\renewcommand{\subsection}{\@startsection{subsection}{2}{\z@}{-3.25ex
plus -1ex minus
   -.2ex}{1.5ex plus .2ex}{\centering\large\sc }}

\makeatletter\renewcommand{\subsubsection}{\@startsection{subsubsection}{3}{-2.45ex}{-3.25ex
plus -1ex minus -.2ex}{1.5ex plus .2ex}{\it }}

\makeatletter\renewcommand{\paragraph}{\@startsection{paragraph}{4}{\z@}%
                                    {0.8ex \@plus1ex \@minus.2ex}%
                                    {-.5em}%
                                    {\normalfont\normalsize\bfseries}}

\renewcommand{\thesection}{\arabic{section}.}
\renewcommand{\thesubsection}{\arabic{section}.\arabic{subsection}.}

\numberwithin{paragraph}{section}
\renewcommand\theparagraph {\S\thesection\@arabic\c@paragraph.\kern-8pt}

\numberwithin{equation}{section}
\renewcommand{\theequation}{\thesection\arabic{equation}}

\renewcommand*\l@section{\@dottedtocline{1}{0em}{1.5em}}
\renewcommand*\l@subsubsection{\@dottedtocline{4}{3.8em}{3.2em}}

\renewcommand\tableofcontents{%
    \section*{\large\contentsname
        \@mkboth{%
           \MakeUppercase\contentsname}{\MakeUppercase\contentsname}}%
       {\baselineskip=15pt plus 2pt minus 1pt
    \@starttoc{toc}}%
\vspace{-3mm}\centerline{{\vrule height 0.5pt width 15.5cm depth
0pt}} }

\renewenvironment{thebibliography}[1]
     {\section*{\centering{\refname}
        \@mkboth{\MakeUppercase\refname}{\MakeUppercase\refname}}%
     \list{\@biblabel{\@arabic\c@enumiv}}%
           {\settowidth\labelwidth{\@biblabel{#1}}%
            \leftmargin\labelwidth
            \advance\leftmargin\labelsep
            \@openbib@code
            \usecounter{enumiv}%
            \let\p@enumiv\@empty
            \renewcommand\theenumiv{\@arabic\c@enumiv}}%
      \sloppy
      \clubpenalty4000
      \@clubpenalty \clubpenalty
      \widowpenalty4000%
      \sfcode`\.\@m}

\DeclareFontFamily{U}{rsf}{}
\DeclareFontShape{U}{rsf}{m}{n}{
  <5> <6> rsfs5 <7> <8> <9> rsfs7 <10-> rsfs10}{}
\DeclareMathAlphabet\Scr{U}{rsf}{m}{n}

\newcommand{\IR}{\mathbbm{R}}    
\newcommand{\IC}{\mathbbm{C}}    
\newcommand{\IZ}{\mathbbm{Z}}    
\newcommand{\IH}{\mathbbm{H}} 
\newcommand{\IP}{\mathbbm{P}}

\newcommand{\CE}{\Scr{E}}
\newcommand{\CF}{\Scr{F}}    
\newcommand{\CH}{\Scr{H}}
    
\newcommand{\CM}{\Scr{M}}    
\newcommand{\CN}{\mathcal{N}} 
\newcommand{\CO}{\Scr{O}} 
\newcommand{\CP}{\Scr{P}}
\newcommand{\CR}{\Scr{R}}
\newcommand{\CS}{\Scr{S}} 
\newcommand{\CU}{\Scr{U}} 

\newcommand{\cN}{\Scr{N}}
\newcommand{\cL}{\Scr{L}}

\renewcommand{\a}{\alpha}
\newcommand{\ad}{{\dot\a}}
\renewcommand{\b}{\beta}
\newcommand{\bd}{{\dot\b}}
\newcommand{\g}{\gamma}
\newcommand{\gd}{{\dot\g}}
\renewcommand{\d}{\delta}
\newcommand{\dd}{{\dot\d}}

\newcommand{\dt}{{\rm d}}

\newcommand{\ua}{{\bf A}}
\newcommand{\ub}{{\bf B}}
\newcommand{\uc}{{\bf C}}
\newcommand{\ud}{{\bf D}}
\newcommand{\ue}{{\bf E}}

\newcommand{\um}{{\bf M}}
\newcommand{\un}{{\bf N}}

\newcommand{\der}[1]{\frac{\partial}{\partial #1}}

\newtheorem{Def}{Definition}[section]
\numberwithin{Def}{section}

\newtheorem{Pro}{Proposition}[section]
\numberwithin{Pro}{section}

\newtheorem{Thm}{Theorem}[section]
\numberwithin{Thm}{section}

\numberwithin{Cor}{section}

\begin{document}

\begin{titlepage}

\setcounter{page}{0}
\renewcommand{\thefootnote}{\fnsymbol{footnote}}

\begin{flushright}
 Imperial--TP--MW--01/07
\end{flushright}

\vspace*{1cm}

\begin{center}

{\LARGE\bf Self-Dual Supergravity and Twistor Theory}

\vspace*{1cm} 

{\Large Martin Wolf} \footnote{E-mail: {\ttfamily m.wolf@imperial.ac.uk}}

\vspace*{1cm}

{\it Theoretical Physics Group\\ 
The Blackett Laboratory, Imperial College London\\
Prince Consort Road\\ London SW7 2AZ, United Kingdom}

\vspace*{1cm}

{\bf Abstract}

\end{center}

\begin{quote}
By generalizing and extending some of the earlier results 
derived by Manin and Merkulov, 
a twistor description is given of four-dimensional
$\CN$-extended (gauged) self-dual supergravity with
and without cosmological constant.
Starting from the category of $(4|4\CN)$-dimensional complex
superconformal supermanifolds, the categories of 
$(4|2\CN)$-dimensional complex quaternionic, 
quaternionic K\"ahler and 
hyper-K\"ahler right-chiral supermanifolds are introduced
and discussed. We then present a detailed twistor description
of these types of supermanifolds. In particular, we construct
supertwistor spaces associated with complex quaternionic 
right-chiral supermanifolds, and explain what additional
supertwistor data allows for giving those supermanifolds a
hyper-K\"ahler structure. In this way, we obtain a
supersymmetric generalization of Penrose's nonlinear graviton
construction. We furthermore give an alternative formulation
in terms of a supersymmetric
extension of LeBrun's Einstein bundle. This allows us to
include the cases with nonvanishing cosmological 
constant. We also discuss the
bundle of local supertwistors and address certain implications
thereof. Finally, we comment on a real version of the theory
related to Euclidean signature. 

\vfill\noindent May 10, 2007

\end{quote}

\setcounter{footnote}{0}\renewcommand{\thefootnote}{\arabic{thefootnote}}

\end{titlepage}

\section{\kern-10pt Introduction and results}

Since the discovery of twistor string theories by Witten
\cite{Witten:2003nn} and Berkovits \cite{Berkovits:2004hg}
about three years ago, 
a lot of advancements in our understanding of the properties of
(supersymmetric) Yang-Mills theory has been made. Despite
the fact that these string theories describe supersymmetric 
Yang-Mills theory coupled to conformal supergravity 
\cite{Berkovits:2004jj},
they provide an elegant way of describing 
some of the remarkable features exhibited by the scattering
amplitudes of the gauge theory (see e.g. \cite{webpage,Cachazo:2005ga}
and references therein). Surely, the appearance of
conformal supergravity is awkward since
it is inextricably mixed in with the gauge theory beyond tree-level
in perturbation theory. This makes it
impossible to solely compute gauge theory scattering amplitudes beyond
tree-level by performing a string theory calculation. In addition,
one rather wishes to describe Einstein supergravity than
conformal supergravity as the latter is believed not to be a suitable 
candidate for describing nature 
due to its lack of unitarity. In view of that,
 Abou-Zeid et al. \cite{Abou-Zeid:2006wu}
proposed new twistor string theories
which indeed seem to yield supersymmetric Yang-Mills theory 
coupled to Einstein supergravity. Among 
the already mentioned aspects, a variety of other related issues has
been investigated and is still being explored
\cite{Neitzke:2004pf}--\cite{Abe:2004ep} (for recent reviews,
see also Refs.\ \cite{Saemann:2006tt}).

Despite the success, a consistent twistor string formulation of gravity 
remains an open question.
In order to find such a formulation, it is
certainly necessary 
to first understand better the twistor description of Einstein
supergravity theories. Before trying to attempt to 
solve this task in full
generality, one may first consider a simplification of
the theory by restricting the focus to the much simpler theory of
self-dual supergravity. In view of that,
recall from the early work by Penrose \cite{Penrose:1976js}
that it is possible to
associate with any complex-Riemannian four-dimensional 
manifold $M$ (complex space-time)
which is equipped with a conformal structure and
has self-dual Weyl curvature, a 
complex three-dimensional twistor space $P$ which is defined to be
the space of maximal isotropic (totally null) complex submanifolds of
$M$.
All the information about the conformal structure of $M$ is
encoded in the complex structure of the twistor space $P$. Some
additional data on $P$ then
allows for the construction of self-dual metrics and 
conformal structures on $M$. For explicit constructions, see
Refs.\ \cite{Ward:1978}--\cite{LeBrun:2005qf},
for instance. Moreover, hidden symmetries and
hierarchies of self-dual gravity have been studied by the
authors of \cite{BoyerAJ}--\cite{Dunajski:2000iq}.
Notice also that one may return to the
realm of Riemannian geometry by restricting the objects
under consideration to the fixed-point set of an
anti-holomorphic involution.

Self-dual supergravity theories on four-dimensional space-time
have first appeared in the works \cite{Kallosh:1979au}--\cite{Siegel:1992wd}
and have subsequently been discussed, e.g. by the authors
of \cite{Devchand:1994hg,Karnas:1997it} 
within the harmonic superspace framework (see also Galperin et
al. \cite{Galperin:2001uw} and references therein). 
The purpose of this article is to give 
the twistor description of
$\CN$-extended self-dual supergravity with and without
cosmological constant. In particular,
we shall generalize and extend the earlier results by Manin \cite{Manin}
and Merkulov \cite{Merkulov:1991kt}--\cite{Merkulov:1992}.\footnote{Notice
that Merkulov \cite{Merkulov:1992qa}
has given a twistor description
of minimal Einstein supergravity.} For most of the time, we work in the 
context of 
complex supermanifolds but at the end we 
also discuss a real version of the theory.
In the next section, starting from
the category of complex superconformal supermanifolds
of dimension $(4|4\CN)$, the categories of
\begin{itemize}
\setlength{\itemsep}{-1mm}
\item[(i)] complex quaternionic right-chiral (hereafter RC) supermanifolds, 
\item[(ii)] complex quaternionic K\"ahler RC supermanifolds and 
\item[(iii)] complex hyper-K\"ahler RC supermanifolds 
\end{itemize}
are introduced and discussed. In this section, special attention is paid to the
construction of the connections and their properties under superconformal
rescalings. In Sec.~\ref{sec:TT}, we first discuss the twistor theory of complex
quaternionic RC supermanifolds. We shall establish a
double fibration of the form 
$$
\begin{aligned}
\begin{picture}(50,40)
\put(0.0,0.0){\makebox(0,0)[c]{$\CP$}}
\put(64.0,0.0){\makebox(0,0)[c]{$\CM$}}
\put(34.0,33.0){\makebox(0,0)[c]{$\CF$}}
\put(7.0,18.0){\makebox(0,0)[c]{$\pi_2$}}
\put(55.0,18.0){\makebox(0,0)[c]{$\pi_1$}}
\put(25.0,25.0){\vector(-1,-1){18}}
\put(37.0,25.0){\vector(1,-1){18}}
\end{picture}
\end{aligned}
$$
where $\CM$ is a complex quaternionic RC supermanifold subject to 
additional restrictions and
$\CP$ its associated supertwistor space. The supermanifold $\CF$ is
a certain $\IP^1$-bundle over $\CM$ and termed correspondence space.
In this way, $\CM$ is viewed as the space of complex submanifolds
of $\CP$ which are biholomorphically equivalent to the complex projective 
line $\IP^1$ and have normal sheaf described by
 $$0\ \longrightarrow\ \Pi\CO_{\IP^1}(1)\otimes\IC^\CN\ \longrightarrow\  
                  \cN_{\IP^1|\CP}\ \longrightarrow\ 
  \CO_{\IP^1}(1)\otimes\IC^2\ \longrightarrow\ 0.$$
Here, $\Pi$ is the Gra{\ss}mann parity changing functor 
and $\CO_{\IP^1}(1)$ is the sheaf of sections of the
dual tautological ($c_1=1$) bundle over $\IP^1$. 
 
Having established this correspondence, we focus on the twistor
description of complex hyper-K\"ahler RC supermanifolds -- the case of 
interest in view of studying self-dual
supergravity with zero cosmological constant.
In particular, we give the supersymmetric analog of 
Penrose's nonlinear graviton construction \cite{Penrose:1976js}, i.e. we shall
show that in this case the supertwistor space is 
holomorphically fibred over the Riemann sphere $\CP\to\IP^1$ and
equipped with a certain relative symplectic structure. 
 Furthermore, we present an equivalent 
formulation of the Penrose construction 
in terms of a supersymmetric generalization of LeBrun's 
Einstein bundle \cite{LeBrun:1985}. This construction allows
for including a cosmological constant to the self-dual
supergravity equations. In particular,
the Einstein bundle is defined over the supertwistor space
and as we shall see, its nonvanishing sections are in one-to-one
correspondence with solutions to the self-dual supergravity equations
with cosmological constant. Requiring its sections to be
integrable amounts to putting the cosmological constant to zero. 
As in the purely bosonic situation, this bundle can explicitly be
described in terms of certain intrinsic holomorphic data on the 
supertwistor space.
Besides this, 
also in Sec.~\ref{sec:TT}, we introduce the bundle of local supertwistors
over $\CM$ and discuss certain implications thereof. For instance, 
we shall show that
it can be reinterpreted in terms of a certain jet-bundle over the supertwistor
space by means of the Penrose-Ward transform. 
All these considerations are first 
generic in the sense of keeping arbitrary the number $\CN$ of allowed 
supersymmetries.
However, like in the flat situation
(see e.g. Witten \cite{Witten:2003nn}), the $\CN=4$ case is special
and deserves a separate treatment.

Finally,
in Sec.~\ref{sec:RS} we discuss a real version of the theory, that is,
we introduce certain real structures (anti-holomorphic involutions)
on all the manifolds appearing
in the above double fibration such that the underlying (ordinary) manifold
of $\CM$ is of Euclidean signature. A particular feature of Euclidean
signature is that the number of allowed supersymmetries is restricted to
be even.

\begin{center}{\sc Remarks}\end{center}

Some general remarks are in order. A {\it complex supermanifold} of 
dimension $m|n$ is meant to be a ringed space $(\CM,\CO_\CM)$, where $\CM$ is a
topological space and $\CO_\CM$ is a sheaf of 
supercommutative ($\IZ_2$-graded) rings on $\CM$ such that, if we let $\cN$ 
be the ideal subsheaf in $\CO_\CM$ of all nilpotent elements, the following is 
fulfilled: 
\begin{itemize}
\setlength{\itemsep}{-1mm}
\item[(i)] $\CM_{\rm red}:=(\CM,\CO_{\rm red}:=\CO_\CM/\cN)$
is a complex manifold of (complex) dimension $m$ and 
\item[(ii)] for any point $x\in\CM$ there is 
neighborhood $\CU\ni x$ such that $\CO_\CM|_\CU\cong\CO_{\rm red}
(\Lambda^\bullet\CE)|_\CU$, 
\end{itemize}
where $\CE:=\cN/\cN^2$ is a rank-$n$ locally free sheaf of
$\CO_{\rm red}$-modules on $\CM$ and $\Lambda^\bullet$
denotes the exterior algebra. We call $\CE$ the {\it characteristic 
sheaf} of the complex supermanifold $(\CM,\CO_\CM)$ and $\CO_\CM$ its 
{\it structure sheaf}.
See, e.g. Manin \cite{Manin} for more
details. Such supermanifolds are said to be
{\it locally split}. In this work, we will 
assume that the Gra{\ss}mann odd directions have trivial topology, that is,
we work in the category of {\it globally split} supermanifolds
$(\CM,\CO_\CM)$ with 
$\CO_\CM\cong\CO_{\rm red}(\Lambda^\bullet\CE)$.
For the sake of brevity, we shall be referring to them as split supermanifolds,
in the sequel. The structure sheaf $\CO_\CM$ of a split supermanifold admits
the $\IZ$-grading $\CO_\CM\cong\bigoplus_{p\geq0}\CO^p_\CM$, where
$\CO^p_\CM\cong\CO_{\rm red}(\Lambda^p\CE)$.
Moreover, the assumption of being split implies that there will always exist an
atlas $\{\{\CU_a\},\{\varphi_{ab},\vartheta_{ab}\}\}$ on $(\CM,\CO_\CM)$ 
such that, if we let $(z_a)=(z^1_a,\ldots,z^m_a)$ be Gra{\ss}mann even coordinates 
and $(\eta_a)=(\eta_a^1,\ldots,\eta_a^n)$ be Gra{\ss}mann odd coordinates on the
patch 
$\CU_a\subset\CM$, the transition functions on nonempty intersections
$\CU_a\cap\CU_b$ are of the form
$z_a=\varphi_{ab}(z_b)$ and $\eta_a=(\vartheta_{j\,ab}^{i}(z_b)\eta_b^j)$
for $i,j=1,\ldots,n$.
We will frequently be working with such atlases without particularly
referring to them. 

An important example of a split supermanifold is the
complex projective superspace $\IP^{m|n}$ given by
$$\IP^{m|n}\ =\ (\IP^m,\CO_{\rm red}(\Lambda^\bullet(\CO_{\IP^m}(-1)
\otimes\IC^n))),$$ where 
$\CO_{\IP^m}(-1)$ is the sheaf of sections of the tautological ($c_1=-1$)
line bundle over the complex projective space $\IP^m$. The reason for the 
appearance of $\CO_{\IP^m}(-1)$ is as follows. If we let 
$(z^0,\ldots,z^m,\eta^1,\ldots,\eta^n)$
be homogeneous coordinates\footnote{Recall that they are
subject to the identification $(z^0,\ldots,z^m,\eta^1,\ldots,\eta^n)\sim 
(tz^0,\ldots,tz^m,t\eta^1,\ldots,t\eta^n)$, where $t\in\IC\setminus\{0\}$.}
 on $\IP^{m|n}$, a holomorphic function $f$ on
$\IP^{m|n}$ has the expansion
$$ f\ =\ \sum f_{i_1\cdots i_r}(z^0,\ldots,z^m)\,\eta^{i_1}\cdots\eta^{i_r}. $$
Surely, for $f$ to be
well-defined the total homogeneity of $f$ must be zero. Hence, 
$f_{i_1\cdots i_r}$ must
be of homogeneity $-r$. This explains the above form of the structure
sheaf of the complex projective superspace.
Of particular interest in the context of (flat) twistor theory 
is $\IP^{3|\CN}$, which is the 
{\it supertwistor space} associated with the superconformal 
compactification of the chiral
complex superspace $\IC^{4|2\CN}$.

Moreover, two things are worth mentioning.
First, {\it any} given complex supermanifold is actually a deformation of 
a split supermanifold (Rothstein \cite{Rothstein:1985}),
that is, to any complex supermanifold $(\CM,\CO_\CM)$ there is 
associated a complex analytic one-parameter family of complex
supermanifolds $(\CM,\CO_{\CM,\, t})$, for $t\in\IC$, such that
$\CO_{\CM,\,t=0}\cong\CO_{\rm red}(\Lambda^\bullet\CE)$ and 
$\CO_{\CM,\,t=1}\cong\CO_\CM$,
where $\CE$ is the characteristic sheaf of $(\CM,\CO_\CM)$.
Second, smooth
supermanifolds are {\it always} split due to Batchelor's theorem
\cite{Batchelor:1979}. The latter result
follows because of the existence of a
smooth partition of unity in the category of smooth supermanifolds.
Since we are eventually interested in self-dual supergravity on
a four-dimensional Riemannian manifold,
this explains why we may restrict our discussion to split supermanifolds
as stated above.

Furthermore, when there is no confusion with the underlying topological
space, we denote the supermanifold $(\CM,\CO_\CM)$ simply by $\CM$.
Finally, we point out that (holomorphic) vector bundles $\CE$ of rank $r|s$ over 
some complex supermanifold 
$(\CM,\CO_\CM)$ are meant to be locally free sheaves of $\CO_\CM$-modules,
that is, they are locally of the form 
$\CO_\CM\otimes\IC^r\oplus\Pi\CO_\CM\otimes\IC^s$.
Hence, the notions
``vector bundle" and ``locally free sheaf" are used interchangeably. 
This will also allow us to simplify notation. In addition, the
dual of any locally free sheaf $\CE$ on $\CM$ is denoted by
$$ \CE^\vee\ =\ \Scr{H}om_{\CO_\CM}(\CE,\CO_\CM). $$
For line bundles $\Scr{L}$, we instead write $\Scr{L}^{-1}$. 
If there is no confusion, the dimensionality
(respectively, the rank) of ordinary manifolds (respectively, of ordinary vector
bundles) will often be
abbreviated by $m|0\equiv m$ (respectively, by $r|0\equiv r$).

\section{\kern -10pt Self-dual supergravity}\label{sec:SDSG}

\subsection{\kern -10pt Superconformal structures}\label{subsec:para}

Remember that a (holomorphic)
{\it conformal structure\/} on an ordinary four-dimensional 
complex spin
manifold $M$ can be introduced in two equivalent ways. The first 
definition states that a conformal structure is an equivalence
class $[g]$, the {\it conformal class}, of holomorphic metrics $g$ on $M$, 
where two given metrics $g$
and $g'$ are called equivalent if $g'=\gamma^2 g$ for some nowhere vanishing
holomorphic function $\gamma$. Putting it differently, a conformal
structure is a line subbundle $L$ in $\Omega^1M\odot \Omega^1M$.
The second definition assumes a
factorization of the holomorphic tangent bundle $TM$ of $M$
as a tensor
product of two rank-2 holomorphic vector 
bundles $S$ and $\tilde S$, that is, 
$TM\cong S\otimes\tilde S$. This isomorphism in turn gives 
(canonically)  
the line subbundle $\Lambda^2S^\vee\otimes\Lambda^2\tilde S^\vee$ in
$\Omega^1M\odot \Omega^1M$ which, in fact, can be identified with
$L$. 

Next one needs to extend the notion of a conformal structure to 
supermanifolds. We shall see that the generalization of the 
latter of the two 
approaches given above seems to be the appropriate one for 
our present purposes (see \ref{rmk:full}~~~for some remarks
regarding standard supergravities). 
Our subsequent discussion 
closely follows the one given by Manin \cite{Manin} and 
Merkulov \cite{Merkulov:1991kt}, respectively.

\paragraph{Superconformal supermanifolds.}
To give the definition of a
{\it superconformal structure}, let $(\CM,\CO_\CM)$ be a $(4|4\CN)$-dimensional
complex supermanifold, where $\CN$ is a nonnegative integer.

{\Def\label{def:para} 
A superconformal structure on $\CM$ is a pair of
integrable rank-$0|2\CN$ distributions $T_l\CM$ and $T_r\CM$ 
which obey the following conditions:
\begin{itemize}
\setlength{\itemsep}{-1mm}
\item[(i)] their sum in the holomorphic tangent bundle $T\CM$ is direct,
\item[(ii)] there exist two rank-$2|0$ locally free sheaves $\CS$ and $\widetilde\CS$ 
and one rank-$0|\CN$ locally free sheaf $\CE$ such that 
$T_l\CM\cong\CS\otimes\CE^\vee$ and $T_r\CM\cong\CE\otimes\widetilde\CS$,\footnote{Do 
not confuse $\CE$ with the characteristic sheaf of $(\CM,\CO_\CM)$.}
\item[(iii)] the Frobenius form
$$
\begin{aligned}
 \Phi\,:\,T_l\CM\otimes T_r\CM\ &\to\ T_0\CM\ :=\ T\CM/(T_l\CM\oplus T_r\CM),\\
 (X\otimes Y)\ &\mapsto\ [X,Y\}\ {\rm mod}\ (T_l\CM\oplus T_r\CM)
\end{aligned}
$$
coincides with the natural map 
$\CS\otimes\CE^\vee\otimes\CE\otimes\widetilde\CS\to\CS\otimes\widetilde\CS$
and gives an isomorphism $T_0\CM\cong\CS\otimes\widetilde\CS$.
Here, $[\cdot,\cdot\}$ denotes the graded Lie bracket.
\end{itemize}
}\vspace*{10pt}

\noindent
From this definition it follows that $T_{l,r}\CM$ define two foliations on 
$\CM$. Let us denote the resulting quotients by $\CM_{r,l}$. Furthermore, 
$\CM_l$ and $\CM_r$ are
supermanifolds which are both of dimension $(4|2\CN)$. In
addition, their structure sheaves $\CO_{\CM_{l,r}}$ are those subsheaves of 
$\CO_\CM$ which are annihilated by vector fields from $T_{r,l}\CM$. By virtue
of the inclusions $\CO_{\CM_{l,r}}\subset\CO_\CM$, we find the following 
double fibration:
\begin{equation}\label{eq:DF1}
\begin{aligned}
\begin{picture}(50,40)
\put(0.0,0.0){\makebox(0,0)[c]{$\CM_l$}}
\put(64.0,0.0){\makebox(0,0)[c]{$\CM_r$}}
\put(34.0,33.0){\makebox(0,0)[c]{$\CM$}}
\put(7.0,18.0){\makebox(0,0)[c]{$\pi_l$}}
\put(55.0,18.0){\makebox(0,0)[c]{$\pi_r$}}
\put(25.0,25.0){\vector(-1,-1){18}}
\put(37.0,25.0){\vector(1,-1){18}}
\end{picture}
\end{aligned}
\end{equation}
According to Manin \cite{Manin}, 
we shall call $\CM_l$ and $\CM_r$ {\it left-} and {\it right-chiral} 
supermanifolds,
respectively. Moreover, we have
\begin{equation}
 0\ \longrightarrow\ T_{l,r}\CM\ \longrightarrow\  
   T\CM\ \longrightarrow\ \pi_{r,l}^*T\CM_{r,l}\ \longrightarrow\ 0,
\end{equation}
which is induced by the double fibration \eqref{eq:DF1}. 
Putting it differently, 
$T_l\CM$ (respectively, $T_r\CM$) is the relative tangent sheaf of the
fibration $\pi_r\,:\,\CM\to\CM_r$ (respectively, of $\pi_l\,:\,\CM\to\CM_l$).
Note that a superconformal structure is, by no means,
just given by a conformal class of supermetrics. 

\paragraph{Some properties and an example.}\label{par:sp}
First of all, it should be noticed that on the underlying four-dimensional
manifold $\CM_{\rm red}$, we naturally have the rank-$2$ holomorphic 
vector bundles $\CS_{\rm red}$ 
and $\widetilde\CS_{\rm red}$. Part (iii) of Def.\
\ref{def:para} then guarantees a factorization of the holomorphic
tangent bundle $T\CM_{\rm red}$ of $\CM_{\rm red}$ as  
$T\CM_{\rm red}\cong \CS_{\rm red}\otimes\widetilde\CS_{\rm red}$. Hence, 
$\CM_{\rm red}$ comes
naturally equipped with a conformal structure.

Furthermore,
Def.\ \ref{def:para} implies that the holomorphic tangent bundle $T\CM$ of $\CM$
fits into the following short exact sequence:
\begin{equation}\label{eq:tanbun}
 0\ \longrightarrow\ T_l\CM\oplus T_r\CM\ 
\overset{i_l\oplus i_r}{\longrightarrow}\  
   T\CM\ \overset{\hat\Phi^{-1}}\longrightarrow\ T_0\CM \longrightarrow\ 0,
\end{equation}
where $i_{l,r}$ are the natural inclusion mappings and $\hat\Phi$ is the 
contracted Frobenius
form which is invertible by assumption. Recall that $T_0\CM\cong \CS\otimes\widetilde\CS$.
Consider now the subsheaves $\pi_{l,r}^*T\CM_{l,r}\subset T\CM$.
It can be shown (cf. Manin \cite{Manin} and Merkulov \cite{Merkulov:1991kt}) that their
structure is also described by a short exact sequence similar
to the one given above, i.e.
\begin{equation}\label{eq:aseq}
 0\ \longrightarrow\ T_{l,r}\CM\ \longrightarrow\  
   \pi^*_{l,r}T\CM_{l,r}\ \longrightarrow\ T_0\CM\ \longrightarrow\ 0.
\end{equation}
This implies that also the underlying manifolds $\CM_{l,r\,{\rm red}}$ of
$\CM_{l,r}$ are naturally equipped with conformal structures in the usual
sense.

The prime example of the above construction is the flag supermanifold
\begin{equation}
\CM\ =\ F_{2|0,2|\CN}(\IC^{4|\CN})\ =\ \{S^{2|0}\subset S^{2|\CN}\subset\IC^{4|\CN}\}.
\end{equation}
In this case, the double
fibration \eqref{eq:DF1} takes the following form:
\begin{equation}\label{eq:DF1-ex}
\begin{aligned}
\begin{picture}(50,40)
\put(-15.0,0.0){\makebox(0,0)[c]{$\CM_l=F_{2|\CN}(\IC^{4|\CN})$}}
\put(74.0,0.0){\makebox(0,0)[c]{$\CM_r=F_{2|0}(\IC^{4|\CN})$}}
\put(34.0,33.0){\makebox(0,0)[c]{$\CM=F_{2|0,2|\CN}(\IC^{4|\CN})$}}
\put(7.0,18.0){\makebox(0,0)[c]{$\pi_l$}}
\put(55.0,18.0){\makebox(0,0)[c]{$\pi_r$}}
\put(25.0,25.0){\vector(-1,-1){18}}
\put(37.0,25.0){\vector(1,-1){18}}
\end{picture}
\end{aligned}
\end{equation}
In addition, there are four natural sheaves $\CS^{2|0}\subset\CS^{2|\CN}$
and $\widetilde\CS^{2|0}\subset\widetilde\CS^{2|\CN}$ on $\CM$, where
$\CS^{2|0}$ and $\CS^{2|\CN}$ are the two tautological sheaves while
the other two are defined by two short exact sequences
\begin{equation}
\begin{aligned}
 0\ \longrightarrow\ \CS^{2|0}\ \longrightarrow\  
   \CO_\CM\otimes\IC^{4|\CN}\ \longrightarrow\ (\widetilde\CS^{2|\CN})^\vee\ 
\longrightarrow\ 0,\\
0\ \longrightarrow\ \CS^{2|\CN}\ \longrightarrow\  
   \CO_\CM\otimes\IC^{4|\CN}\ \longrightarrow\ (\widetilde\CS^{2|0})^\vee\ 
\longrightarrow\ 0.
\end{aligned}
\end{equation}
A short calculation shows that
these two sequences together with \eqref{eq:aseq} imply 
\begin{equation}
 \CS\ \cong\ (\CS^{2|0})^\vee,\qquad
 \widetilde\CS\ \cong\ (\widetilde\CS^{2|0})^\vee\qquad{\rm and}\qquad
 \CE\ \cong\ \widetilde\CS^{2|\CN}/\widetilde\CS^{2|0}.
\end{equation}
In addition, one may also verify that points (i) and (iii) of Def.~\ref{def:para}
are satisfied. For more details, see Manin \cite{Manin}.
The three flag supermanifolds
$F_{2|0,2|\CN}(\IC^{4|\CN})$, $F_{2|\CN}(\IC^{4|\CN})$ and
$F_{2|0}(\IC^{4|\CN})$ play an important role in the twistor description
of supersymmetric Yang-Mills theories, as they represent the superconformal
compactifications of the flat complex superspaces $\IC^{4|4\CN}$ and 
$\IC_{l,r}^{4|2\CN}$
(for recent reviews, see Refs.~\cite{Saemann:2006tt}).

\paragraph{Remarks.}\label{rmk:full}
In this work, we shall only be concerned with 
$\CN$-extended self-dual supergravities. 
These theories are conveniently formulated on
right-chiral supermanifolds, as has been discussed in, e.g., 
Refs.~\cite{Siegel:1992wd}--\cite{Karnas:1997it}. By
a slight abuse of notation, we denote $\CM_r$ simply by $\CM$. Henceforth,
we shall be working with (complex) 
right-chiral (hereafter RC) supermanifolds of dimension $(4|2\CN)$. Furthermore,
for various technical reasons but also for reasons related to the real structures 
discussed in Sec.~\ref{sec:RS}, we restrict the 
number $\CN$ of allowed supersymmetries to be {\it even\/}. 
Note that for the complex situation (or the case of split signature),
this restriction on $\CN$ is not necessary.

Furthermore, let us emphasize 
that for making contact with standard {\it full\/} non-self-dual supergravities 
with $\CN>2$ (this includes both Einstein and conformal supergravities), 
a superconformal structure should be defined slightly 
differently since due to certain torsion constraints (see e.g. 
Howe \cite{howe}), there are no (anti-)chiral superfields. This means that
the distributions $T_{l,r}\CM$ should be taken to be suitably
non-integrable, which means that one does not have the double fibration
\eqref{eq:DF1}. Also for $\CN>2$, one should relax the condition
of integrability of $T_l\CM$ if one wishes to talk about self-dual
supergravity since in that case only $T_r\CM$ should be considered
to be integrable. For our discussions given
below, these issues can be left aside. They certainly
deserve further studies in view of complementing the self-dual theory
to the full theory.

\subsection{\kern-10pt Geometry of right-chiral supermanifolds}\label{sec:pre}

This section is devoted to some geometric aspects of RC supermanifolds.
In particular, we discuss their structure group, introduce scales and
vielbeins and talk about connections, torsion and curvature.

\paragraph{Structure group.}\label{par:SG}
Let $\CM$ be an RC supermanifold.
From \eqref{eq:aseq} we know that the holomorphic tangent bundle
$T\CM$ of $\CM$ is described by a sequence of the form
\begin{equation}\label{eq:ES1}
 0\ \longrightarrow\ \CE\otimes\widetilde\CS\ \longrightarrow\  
   T\CM\ \longrightarrow\ \CS\otimes\widetilde\CS\ \longrightarrow\ 0,
\end{equation}
where $\CS$ and $\widetilde\CS$ are both of rank $2|0$
and $\CE$ is of rank
$0|\CN$, respectively. 
By a slight abuse of notation, we are again using 
the same symbols $\CS$, $\widetilde\CS$ and $\CE$.
Next we notice that 
$T\CM$ is given by the tensor product $\CH\otimes\widetilde\CS$,
where $\CH\to\CM$ is a rank-$2|\CN$ holomorphic vector bundle over $\CM$
described by 
\begin{equation}\label{eq:ES2}
 0\ \longrightarrow\ \CE\ \longrightarrow\  
   \CH\ \longrightarrow\ \CS\ \longrightarrow\ 0.
\end{equation}
Hence, the structure group of $T\CM$ is as follows: the supergroup
$GL(2|\CN,\IC)$ acts on the left on $\CH$ and $GL(2|0,\IC)$
acts on the right on $\widetilde\CS$ by inverses. The resulting induced action
on the tangent bundle $T\CM$ yields a subsupergroup of $GL(4|2\CN,\IC)$.
Let us denote it by $G$ and its Lie superalgebra by $\mathfrak{g}$.
In addition, there is a $|4-\CN|$-fold cover of $G\subset GL(4|2\CN,\IC)$
\begin{equation}\label{eq:cover}
 1\ \longrightarrow\ \IZ_{|4-\CN|}\ \longrightarrow\ S(GL(2|\CN,\IC)\times GL(2|0,\IC))\ 
 \longrightarrow\ G\ \longrightarrow\ 1,
\end{equation}
where $S(GL(2|\CN,\IC)\times GL(2|0,\IC))$ is the subsupergroup 
of $SL(4|\CN,\IC)$ consisting of matrices of the form
\begin{equation}
 \begin{pmatrix} A_1 & 0 & A_2\\ 0 & b & 0\\ A_3 & 0 & A_4\end{pmatrix},\qquad{\rm with}\qquad
 \begin{pmatrix} A_1 & A_2\\ A_3 & A_4\end{pmatrix}\ \in\ GL(2|\CN,\IC)
\end{equation}
and $b\in GL(2|0,\IC)$.
Here, the $A_i$s, for $i=1,\ldots,4$, are the defining blocks of a supermatrix in standard format,
that is, the matrices $A_{1,4}$ are Gra{\ss}mann even while the $A_{2,3}$ are Gra{\ss}mann odd (see e.g.
Manin \cite{Manin} for more 
details).

\paragraph{Scales and vielbeins.}\label{par:SaV}
Let us start by defining what shall be understood by a {\it scale} on an
RC supermanifold $\CM$
(cf. also Refs.\ \cite{Manin,Merkulov:1991kt,Bailey:1991,Merkulov:1992}).
Consider the sequences \eqref{eq:ES1} and \eqref{eq:ES2}. They 
immediately give the following natural isomorphisms of Berezinian
sheaves:
\begin{equation}
{\rm Ber}\,T\CM\ \cong\ ({\rm Ber}\,\CS)^2\otimes({\rm Ber}\,\CE)^2
\otimes({\rm Ber}\,\widetilde\CS)^{2-\CN}
\ \cong\ ({\rm Ber}\,\CH)^2\otimes({\rm Ber}\,\widetilde\CS)^{2-\CN}.
\end{equation}
Hence, the Berezinian sheaf ${\rm Ber}(\CM):={\rm Ber}\,\Omega^1\CM$ of $\CM$ is
\begin{equation}\label{eq:beriso1}
 {\rm Ber}(\CM)\ \cong\ ({\rm Ber}\,\CH^\vee)^2\otimes({\rm Ber}\,\widetilde\CS^\vee)^{2-\CN}.
\end{equation}
In addition, we have ${\rm Ber}\,\widetilde\CS\cong\Lambda^2\widetilde\CS$
since $\widetilde\CS$ is of purely even rank $2|0$. 
Moreover, without loss of generality, one may always make the identification
\begin{equation}\label{eq:beriso}
 {\rm Ber}\,\CH\ \cong\ {\rm Ber}\,\widetilde\CS.
\end{equation}
In this respect, recall that $\CN$ is assumed to be even.
In fact, since the tangent bundle can be factorized as $T\CM\cong\CH\otimes\widetilde\CS$,
one can locally choose to have such an isomorphism. 

Then we may give the following definition:
{\Def A scale on an RC supermanifold $\CM$ is a choice of a particular
non-vanishing volume form $\tilde\varepsilon\in H^0(\CM,{\rm Ber}\,\widetilde\CS^\vee)$
on the vector bundle $\widetilde\CS$. A superconformal rescaling is a change of
scale.
}\vspace*{10pt}

\noindent
Therefore, together with the identification 
${\rm Ber}\,\CH^\vee\cong{\rm Ber}\,\widetilde\CS^\vee$,
a section of ${\rm Ber}\,\widetilde\CS^\vee$ gives a section of ${\rm Ber}(\CM)$
on $\CM$. We shall denote a generic section of ${\rm Ber}(\CM)$
by {\rm Vol}, in the sequel.\footnote{Later on, we additionally require that 
the resulting volume form
${\rm Vol}\in H^0(\CM,{\rm Ber}(\CM))$ obeys
$\rho({\rm Vol})= {\rm Vol}$, where $\rho$ is a real structure on $\CM$.
See Sec.~\ref{sec:RS} for more details.}

It is well known that a
particular choice of a coordinate system on any supermanifold determines
the corresponding trivialization of the (co)tangent bundle and hence, of the
Berezinian bundle. Let now 
$\CU$ be an open subset of $\CM$. On $\CU$ we may introduce  
$(x^{\mu\dot\nu},\eta^{m\dot\mu})$ as local coordinates, where
$\mu,\nu,\ldots=1,2$, $\dot\mu,\dot\nu,\ldots=\dot1,\dot2$ and
$m,n,\ldots=1,\ldots,\CN$. The entire set of coordinates is denoted
by $x^\um$, where $\um=(\mu\dot\nu,m\dot\mu)$ is an Einstein
index. We shall also make use of the notation
$\um=M\dot\mu$, where $M=(\mu,m)$. 
Then $\partial/\partial x^\um$
(respectively, $\dt x^\um$) are basis sections of the tangent
bundle $T\CM$ (respectively, of the cotangent bundle $\Omega^1\CM$)
 of $\CM$.
We may associate with the set $\{\partial/\partial x^\um\}$
(respectively, with $\{\dt x^\um\}$) a basis section of Ber$(\CM)$
which we denote by $D^{-1}(\partial/\partial x^\um)$ (respectively,
by $D(\dt x^\um)$). An arbitrary (local) section of Ber$(\CM)$ then
takes the following form:
\begin{equation}\label{eq:defvol}
 {\rm Vol}\ =\ \phi\, D^{-1}\left(\der{x^\um}\right)\ =\
          \phi\,D(\dt x^\um)\ =:\ \phi\,\dt^4x\,\dt^{2\CN}\eta,
\end{equation}
where $\phi$ is a nonvanishing function on $\CU\subset\CM$. In the
last step in the above equation, 
we have introduced a more conventional notation for the volume form.    

Next we introduce (local) frame fields $E_\ua$, which generate the tangent bundle 
$T\CM$, by setting
\begin{equation}\label{eq:defvol2}
 {\rm Vol}\ =\ D^{-1}(E_\ua),\qquad{\rm with}\qquad E_\ua\ :=\ {E_\ua}^\um\der{x^\um}.
\end{equation}
Obviously, frame fields are unique up to $SG$-transformations of
the form $E_\ua\mapsto {C_\ua}^\ub E_\ub$, where $C=({C_\ua}^\ub)$ is
an $SG$-valued function on $\CU$ with $SG$ being the subsupergroup of
$G$ described by
\begin{equation}\label{eq:cover1}
 1\ \longrightarrow\ \IZ_2\ \longrightarrow\ SL(2|\CN,\IC)\times SL(2|0,\IC)\ 
 \longrightarrow\ SG\ \longrightarrow\ 1,
\end{equation}
as $\CN$ is assumed to be even. Putting it
differently, a choice of scale on $\CM$ reduces the structure group from
$G$ to $SG$.
By comparing \eqref{eq:defvol} with \eqref{eq:defvol2}, we see that  
the function $\phi$ is given by the superdeterminant of ${E_\ua}^\um$, i.e.
$\phi={\rm Ber}({E_\ua}^\um)$. This particular frame is also called
the {\it structure frame}. In 
addition, (local) coframe fields $E^\ua$ are given by
\begin{equation}\label{eq:coframe}
E^\ua\ :=\ \dt x^\um\, {E_\um}^\ua,\qquad{\rm with}\qquad E_\ua\lrcorner E^\ub\ =\ {\d_\ua}^\ub.
\end{equation}
They generate the 
cotangent bundle $\Omega^1\CM$ of $\CM$. Here, ${E_\ua}^\um$ and
${E_\um}^\ua$ are called {\it vielbein matrices} which obey
\begin{equation}
 {E_\ua}^\um {E_\um}^\ub\ =\ {\d_\ua}^\ub\qquad{\rm and}\qquad
 {E_\um}^\ua {E_\ua}^\un\ =\ {\d_\um}^\un.
\end{equation}
The {\it structure group} or {\it structure frame} indices 
$\ua,\ub,\ldots$ look explicitly as
$\ua=(\a\bd,i\ad)$, where 
$\a,\b,\ldots=1,2$, $\ad,\bd,\ldots=\dot1,\dot2$ and
$i,j,\ldots=1,\ldots,\CN$, respectively. Again, we shall
write $\ua=A\ad$ with $A=(\a,i)$.

Recall that by virtue of \eqref{eq:beriso1} and
\eqref{eq:beriso}, a section of ${\rm Ber}\,\widetilde\CS^\vee$
gives a section of ${\rm Ber}(\CM)$. If we rescale this section by some
nonvanishing function $\gamma$, the
volume form Vol changes as ${\rm Vol}\mapsto\widehat{\rm Vol}= \g^{4-\CN}{\rm Vol}$.
Up to $SG$-transformations (which can always be reabsorbed in the definition
of the vielbein),
the frame and coframe fields change accordingly as $E_\ua\mapsto 
E_{\widehat{\ua}}= \g^{-\kappa} E_{\ua}$
and $E^\ua\mapsto E^{\widehat{\ua}}= \g^\kappa E^{\ua}$, where
\begin{equation}\label{eq:defkap}
 \kappa\ :=\ \frac{4-\CN}{4-2\CN}.
\end{equation}

\paragraph{Connection.}\label{par:CTC} 
Generally speaking,
an affine connection $\nabla$ on $\CM$ is a Gra{\ss}mann {\it even} mapping
on the tangent bundle $T\CM$,
\begin{equation}
 \nabla\,:\,T\CM\ \to\ T\CM\otimes\Omega^1\CM,
\end{equation}
which satisfies the Leibniz formula
\begin{equation}\label{eq:defnab}
 \nabla(f X)\ =\ \dt f\otimes X+f\nabla X,
\end{equation}
where $f$ is a local holomorphic function and $X$ a local
section of $T\CM$. Setting $\nabla_\ua:=E_\ua\lrcorner\nabla$, we may
write Eq.
\eqref{eq:defnab} explicitly as
\begin{equation}
 \nabla_\ua(f X)\ =\ (E_\ua f)X+(-)^{p_fp_\ua}f\nabla_\ua X.
\end{equation}
Here, $p\in\IZ_2$ denotes the Gra{\ss}mann parity.
Since $T\CM\cong\CH\otimes\widetilde\CS$, we have the decomposition
\begin{equation}\label{eq:nabdec}
 \nabla\ =\ \nabla_\CH\otimes{\rm id}_{\widetilde\CS}+
 {\rm id}_\CH\otimes\nabla_{\widetilde\CS},
\end{equation}
where 
\begin{equation}
 \nabla_\CH\,:\,\CH\ \to\ \CH\otimes\Omega^1\CM\qquad{\rm and}\qquad
 \nabla_{\widetilde\CS}\,:\,\widetilde\CS\ \to\ \widetilde\CS\otimes\Omega^1\CM
\end{equation}
are the two connections on $\CH$ and $\widetilde\CS$, respectively. 
Locally, the connection $\nabla$ is given in terms of a $\mathfrak{g}$-valued
connection
one-form $\Omega=({\Omega_\ua}^\ub)=
(E^\uc{\Omega_{\uc\ua}}^\ub)$ which
is defined by
\begin{equation}\label{eq:AC}
 \nabla E_\ua\ =\  {\Omega_\ua}^\ub E_\ub,
\end{equation}
with
\begin{equation}
 {\Omega_\ua}^\ub\ =\ {\Omega_{A\ad}}^{B\bd}\ =\ 
 {\Omega_A}^B{\d_\ad}^\bd+{\d_A}^B{\Omega_\ad}^\bd,
\end{equation}
by virtue of \eqref{eq:nabdec}.
Therefore, Eqs.\ \eqref{eq:AC} read explicitly as
$$
\nabla E_\ua\ =\  \nabla E_{A\ad}\ =\ {\Omega_A}^B E_{B\ad}+{\Omega_\ad}^\bd E_{A\bd}.
$$
In the following,
we shall not make any notational distinction between the three connections
$\nabla$, $\nabla_\CH$ and $\nabla_{\widetilde\CS}$ and simply denote them commonly by
$\nabla$. It will be clear from the context which of those is actually being considered.

\paragraph{Torsion.}\label{par:CTC2}
If we set
\begin{equation}
 [E_\ua,E_\ub\}\ =\ {f_{\ua\ub}}^\uc E_\uc,
\end{equation}
where
${f_{\ua\ub}}^\uc$ are the structure functions,
the components of the torsion $T=T^\ua E_\ua=\frac{1}{2}E^\ub\wedge 
E^\ua {T_{\ua\ub}}^\uc E_\uc$ 
of $\nabla$, 
which is defined by
\begin{equation}\label{eq:deftor}
 T^\ua\ =\ -\nabla E^\ua\ =\ -\dt E^\ua+E^\ub\wedge{\Omega_\ub}^\ua,
\end{equation}
are given by
\begin{equation}\label{eq:torsioncomp}
 {T_{\ua\ub}}^\uc\ =\ {\Omega_{\ua\ub}}^\uc-(-)^{p_\ua p_\ub}{\Omega_{\ub\ua}}^\uc
 -{f_{\ua\ub}}^\uc.
\end{equation}
Note that if we consider the space of differential two-forms on $\CM$, we have 
\begin{equation}\label{eq:2formdec}
 \Lambda^2\Omega^1\CM\ \cong\ \Lambda^2(\CH^\vee\otimes\widetilde\CS^\vee)\ \cong\
      (\Lambda^2\CH^\vee\otimes\odot^2\widetilde\CS^\vee)\oplus
      (\odot^2\CH^\vee\otimes \Lambda^2\widetilde\CS^\vee),
\end{equation}
where $\odot^p$ denotes the $p$-th (graded) 
symmetric power of the bundles in question.
Therefore, $T$ can be decomposed as 
\begin{equation}
 T\ =\ T^-+T^+,
\end{equation}
with
\begin{equation}
 T^-\ \in\ H^0(\CM,\Lambda^2\CH^\vee\otimes \odot^2
 \widetilde\CS^\vee\otimes T\CM)\quad{\rm and}\quad
 T^+\ \in\ H^0(\CM,\odot^2\CH^\vee\otimes \Lambda^2
 \widetilde\CS^\vee\otimes T\CM).
\end{equation}
In the structure frame, $T^\mp$ look in components as
\begin{equation}
 T^-\ :\ {T_{A(\ad B\bd)}}^{C\gd}\qquad{\rm and}\qquad
 T^+\ :\ {T_{A[\ad B\bd]}}^{C\gd},
\end{equation}
where parentheses denote normalized symmetrization while square brackets denote
normalized antisymmetrization, respectively. 

A tensor is called {\it totally trace-free}
if {\it all} possible supertraces with respect to upper and lower
indices vanish. Then we have the following proposition:

{\Pro On any RC supermanifold $\CM$ with fixed scale,
the totally trace-free parts of $T^-$ and of $T^+$
are independent of the choice of connection, i.e. they are 
invariants of $\CM$.}\vspace*{10pt} 

\noindent
{\it Proof:} Here, we are following ideas of
Ref.\ \cite{Bailey:1991} but adopted to
the supersymmetric setting.
Recall that 
$T\CM\cong\CH\otimes\widetilde\CS$. 
Let $\mu^A$ be a section of $\CH$ and
$\lambda^\ad$ be a section of $\widetilde\CS$, respectively. 
For fixed scale, a general change of a given connection 
$\nabla_{A\ad}=E_{A\ad}\lrcorner\nabla$ to another one 
$\widehat\nabla_{A\ad}=E_{A\ad}\lrcorner\widehat\nabla$ is given in terms of the 
contorsion tensors
${\Theta_{A\ad B}}^C$ and ${\Theta_{A\ad\bd}}^\gd$ by
$$
 (\widehat\nabla_{A\ad}-\nabla_{A\ad})\mu^B\ =\ 
 (-)^{p_Ap_C}\mu^C{\Theta_{A\ad C}}^B\qquad{\rm and}\qquad
 (\widehat\nabla_{A\ad}-\nabla_{A\ad})\lambda^\bd\ =\ 
 \lambda^\gd{\Theta_{A\ad\gd}}^\bd.
$$
Hence, for a section $u^{A\ad}$ of $T\CM$ this implies
$$
 (\widehat\nabla_{A\ad}-\nabla_{A\ad})u^{B\bd}\ =\ (-)^{p_Ap_C}u^{C\gd} 
 {\Theta_{A\ad C\gd}}^{B\bd},
$$
with
$$ {\Theta_{A\ad B\bd}}^{C\gd}\ =\ {\Theta_{A\ad B}}^C{\d_\bd}^\gd+
 {\d_B}^C{\Theta_{A\ad\bd}}^\gd.
$$
Note that $p_{\ua}\equiv p_A$. From $$[\widehat\nabla_{A\ad},\widehat\nabla_{B\bd}\}f\ =\ 
-{\widehat T}_{A\ad B\bd}\ \!^{\kern-2pt C\gd}\widehat\nabla_{C\gd}f,$$ where $f$ is 
a local section
of $\CO_\CM$, and from similar expressions for unhatted quantities, we thus obtain
\begin{equation}\label{eq:changetors}
\begin{aligned}
 {\widehat T}_{A(\ad B\bd)}\ \!^{\kern-2pt C\gd}\ &=\ {T_{A(\ad B\bd)}}^{C\gd}-
     2{\Theta_{[A(\ad B\}}}^C{\d_{\bd)}}^\gd-2{\Theta_{[A(\ad\bd)}}^\gd{\d_{B\}}}^C,\\
 {\widehat T}_{A[\ad B\bd]}\ \!^{\kern-2pt C\gd}\ &=\ {T_{A[\ad B\bd]}}^{C\gd}-
     2{\Theta_{\{A[\ad B]}}^C{\d_{\bd]}}^\gd-2{\Theta_{\{A[\ad\bd]}}^\gd{\d_{B]}}^C,
\end{aligned}
\end{equation}
where $[\cdot\}$ denotes normalized graded antisymmetrization of the enclosed indices
while $\{\cdot]$ means normalized graded symmetrization. These expressions make it
obvious that changes in the connection are only reflected in the trace parts
of the torsion.

\hfill $\Box$ 

\noindent
Therefore, without loss of generality, we can always work with a connection $\nabla$ on
$\CM$ whose torsion tensors $T^-$ and $T^+$
are totally trace-free, since given any two connections on the bundles $\CH$ and
$\widetilde\CS$ it is always possible to find contorsion tensors such the resulting 
connection induced on the tangent bundle $T\CM$ will be totally trace-free.
 Indeed, we have the following
proposition:

{\Pro\label{pro:LC1} 
On any RC supermanifold $\CM$ 
with fixed scale $\tilde\varepsilon\in H^0(\CM,{\rm Ber}\,\widetilde\CS^\vee)$
there always exits a connection such that: 
\begin{itemize}
\setlength{\itemsep}{-1mm}
\item[(i)] the torsion tensors $T^-$ and $T^+$ are totally trace-free and
\item[(ii)] in addition we have that
$$ \nabla\varepsilon\ =\ 0\ =\ \nabla\tilde\varepsilon, $$
where $\varepsilon\in H^0(\CM,{\rm Ber}\,\CH^\vee)$ is determined by $\tilde\varepsilon$
via the isomorphism \eqref{eq:beriso}.
\end{itemize}
Furthermore, for $\CN\neq4$, this connection is unique.
}\vspace*{10pt}

\noindent
{\it Proof:} Existence is clear from our above discussion. It remains to prove uniqueness
for $\CN\neq4$.
First of all, one notices that given two connections $\nabla$ and $\widetilde\nabla$
whose torsion tensors $T^\mp$ and $\widetilde T^\mp$ are totally trace-free, then
their contorsion tensors $\Theta$ and $\widetilde\Theta$ (obtained from an arbitrary connection
one has started with) can only differ by the following terms:
$$
\begin{aligned}
 {\widehat\Theta_{A\ad B}}\,\!^C\ &:=\ {\widetilde\Theta_{A\ad B}}\,\!^C-{\Theta_{A\ad B}}^C\ =\
 X_{\{A\ad}{\d_{B]}}^C+ 
Y_{[A\ad}{\d_{B\}}}^C,\\
 {\widehat\Theta_{A\ad \bd}}\,\!^\gd\ &:=\ {\widetilde\Theta_{A\ad \bd}}\,\!^\gd-
 {\Theta_{A\ad\bd}}^\gd\ =\ -Y_{A(\ad}{\d_{\bd)}}^\gd-X_{A[\ad}{\d_{\bd]}}^\gd,
\end{aligned}
$$
where $X_{A\ad}$ and $Y_{A\ad}$ are arbitrary differential one-forms on $\CM$.
This can be seen upon inspecting the Eqs.\ \eqref{eq:changetors}.
Next one picks a volume form $\tilde\varepsilon\in H^0(\CM,{\rm Ber}\,\widetilde\CS^\vee)$
and hence a volume form $\varepsilon\in H^0(\CM,{\rm Ber}\,\CH^\vee)$. In a
structure frame, they are of the form (recall that $\CM$ is split)
$$ {\varepsilon_{\a\b}}^{i_1\cdots i_\CN}\ =\ \epsilon_{\a\b}\epsilon^{i_1\cdots i_\CN}
   \qquad{\rm and}\qquad
   \tilde\varepsilon_{\ad\bd}\ =\ \epsilon_{\ad\bd},$$
where the $\epsilon$-tensors are totally antisymmetric with
$\epsilon_{12}=\epsilon_{\dot1\dot2}=-\epsilon^{1\cdots\CN}=-1$. Since $\CN$ is
assumed to be even, we find 
$$
\begin{aligned}
 \widetilde\nabla_{A\ad}{\varepsilon_{\b\g}}^{j_1\cdots j_\CN}\ &=\
 \nabla_{A\ad}{\varepsilon_{\b\g}}^{j_1\cdots j_\CN}-2{\widehat \Theta_{A\ad[\b}}\,\!^\d
 {\varepsilon_{\d\g]}}^{j_1\cdots j_\CN}+\CN{\varepsilon_{\b\g}}^{[j_1\cdots 
   j_{\CN-1}k}{\widehat \Theta_{A\ad k}}\,\!^{j_\CN]},\\
 \widetilde\nabla_{A\ad}\varepsilon_{\bd\gd}\ &=\ \nabla_{A\ad}\varepsilon_{\bd\gd}-2
 {\widehat \Theta_{A\ad[\bd}}\,\!^\dd\varepsilon_{\dd\gd]}.
\end{aligned}
$$
These equations in turn imply that 
$$ (-)^B\widehat{\Theta}_{A\ad B}\,\!^B\ =\ 
 \widehat \Theta_{A\ad\b}\,\!^\b-\widehat \Theta_{A\ad j}\,\!^j\ =\ 0\ =\ 
   \widehat \Theta_{A\ad\bd}\,\!^\bd,
$$
since both, $\nabla$ and $\widetilde\nabla$ are assumed
to annihilate $\varepsilon$ and $\tilde\varepsilon$, respectively.
It is then a rather straightforward exercise to verify that $X_{A\ad}$ and
$Y_{A\ad}$ must vanish for $\CN\neq 4$. Hence, $\widetilde\nabla=\nabla$ and the proof is 
completed. 

\hfill $\Box$ 

\noindent
Henceforth, we shall be working with a connection on $\CM$ 
which has totally trace-free torsion tensors $T^\mp$.
Note that
if 
$T^+$ is taken to be totally trace-free, it must vanish identically.
This is seen as follows.
One first notices that
 ${T_{A[\ad B\bd]}}^{C\gd}=\epsilon_{\ad\bd}{T_{AB}}^{C\gd}$
as the rank of $\widetilde\CS$ is $2|0$.  Since
$T^-$ is totally trace-free, it follows from
$${T_{A[\ad B\bd]}}^{C\bd}\ =\ \epsilon_{\ad\bd}{T_{AB}}^{C\bd}\ =\ 0$$ 
that ${T_{AB}}^{C\gd}=0$. Altogether,
the torsion tensor takes the form
\begin{equation}\label{eq:redtors}
 T\ =\ T^-.
\end{equation}

{\Def\label{def:CQRC}
 An RC supermanifold $\CM$ is said to be complex quaternionic
 it is equipped with a torsion-free connection which annihilates
 both volume forms $\varepsilon$ and $\tilde\varepsilon$. 
}\vspace*{10pt}

\noindent

For our later discussions, we need to know how a connection behaves
under changes of scale. 

{\Pro Suppose we are given an RC supermanifold $\CM$ which is
equipped with a connection $\nabla$ that 
obeys conditions (i) of (ii) given in
 Prop.\ \ref{pro:LC1} Suppose further that $\CN\neq4$.
Under a rescaling of the form
$\tilde\varepsilon\mapsto\g\tilde\varepsilon$, where $\g$ is a nonvanishing
holomorphic function, the change of connection
to the new one $\widehat\nabla$ is given by the following contorsion tensors:
\begin{equation}\label{eq:changecon}
\begin{aligned}
{\Theta_{A\ad B}}^C\ &=\ (-)^{p_Ap_B}\g_{B\ad}{\d_A}^C-
\tfrac{\kappa}{2}\gamma_{A\ad}{\d_B}^C,\\
{\Theta_{A\ad\bd}}^\gd\ &=\ \gamma_{A\bd}{\d_\ad}^\gd-
\tfrac{\kappa}{2}\gamma_{A\ad}{\d_\bd}^\gd. 
\end{aligned}
\end{equation}
Here, $\g_{A\ad}:=E_{A\ad}\log\g$ and the constant 
$\kappa$ has been introduced in \eqref{eq:defkap}. This implies that the
new connection $\widehat\nabla_{\widehat A\widehat\ad}=
   E_{\widehat A\widehat\ad}\lrcorner\widehat\nabla$,
with $E_{\widehat A\widehat\ad}=\g^{-\kappa}E_{A\ad}$,
acts as follows:
\begin{equation}\label{eq:TLCRL}
\begin{aligned}
 \nabla_{A\ad}\mu^B\ &\mapsto\ \widehat
   \nabla_{\widehat A\widehat\ad}\mu^{\widehat B}\kern-7pt &&=\ 
  \g^{-\frac{1}{2}\kappa}(\nabla_{A\ad}\mu^B+{\d_A}^B\mu^C\g_{C\ad}),\\
 \nabla_{A\ad}\lambda^\bd\ &\mapsto\ 
       \widehat\nabla_{\widehat A\widehat\ad}\lambda^{\widehat\bd}\kern-7pt &&=\ 
  \g^{-\frac{1}{2}\kappa}(\nabla_{A\ad}\lambda^\bd+
      {\d_\ad}^\bd\lambda^\gd\g_{A\gd}),\\
 \nabla_{A\ad}\mu_B\ &\mapsto\ 
   \widehat\nabla_{\widehat A\widehat\ad}\mu_{\widehat B}\kern-7pt &&=\
  \g^{-\frac{3}{2}\kappa}(\nabla_{A\ad}\mu_B-\mu_A\gamma_{B\ad}),\\ 
 \nabla_{A\ad}\lambda_\bd\ &\mapsto\ 
   \widehat\nabla_{\widehat A\widehat\ad}\lambda_{\widehat\bd}\kern-7pt &&=\
  \g^{-\frac{3}{2}\kappa}(\nabla_{A\ad}\lambda_\bd-\g_{A\bd}\lambda_\ad),\\ 
\end{aligned}
\end{equation}
where $\mu^A$ and $\lambda^\ad$ 
are sections of the vector bundles $\CH$ and $\widetilde\CS$,
respectively, together with $\mu^{\widehat A}=\g^{\frac{1}{2}\kappa}\mu^A$
and $\lambda^{\widehat\ad}=\g^{\frac{1}{2}\kappa}\lambda^\ad$ and similarly for their duals.
  }\vspace*{10pt}

\noindent
{\it Proof:} The first thing one notices is that the components of
the volume forms ${\varepsilon_{\a\b}}^{i_1\cdots i_\CN}$ and $\tilde\varepsilon_{\ad\bd}$ 
scale as
$$
 {\varepsilon_{\a\b}}^{i_1\cdots i_\CN}\ \mapsto\
 {\widehat\varepsilon_{\widehat\a\widehat\b}}\,\!^{\widehat i_1\cdots\widehat i_\CN}\ =\ 
 \g^{\frac{\CN}{4}}{\varepsilon_{\a\b}}^{i_1\cdots i_\CN}\qquad{\rm and}\qquad
 \tilde\varepsilon_{\ad\bd}\ \mapsto\ \widehat{\tilde\varepsilon}_{\widehat\ad\widehat\bd}\ =\ 
 \g^{-\frac{\CN}{4-2\CN}}\tilde\varepsilon_{\ad\bd}.
$$
Hence, the conditions
$$
 \widehat\nabla_{\widehat A\widehat\ad}\,{\widehat\varepsilon_{\widehat\a\widehat\b}}
  \,\!^{\widehat i_1\cdots\widehat i_\CN}\ =\ 0\ =\ 
  \widehat\nabla_{\widehat A\widehat\ad}\,\widehat{\tilde\varepsilon}_{\widehat\ad\widehat\bd}
$$
yield
$$
 (-)^B\Theta_{A\ad B}\,\!^B\ =\ {\Theta_{A\ad\b}}^\b-{\Theta_{A\ad j}}^j\ =\ 
 \tfrac{\CN}{4}\g_{A\ad}\qquad{\rm and}\qquad
 {\Theta_{A\ad\bd}}^\bd\ =\ -\tfrac{\CN}{4-2\CN}\g_{A\ad}.
$$
Furthermore, by the requirement that the parts $\widehat T^\mp$ of $\widehat T$ 
are totally trace-free, we
find that (see also the proof of Prop.\ \ref{pro:LC1}) 
$$
 {\Theta_{[A\ad B\bd\}}}^{C\gd}\ =\ -\kappa\,\g_{[A\ad}{\d_{B\bd\}}}^{C\gd}.
$$
Combining these results, we arrive after some algebra at Eqs.\ \eqref{eq:changecon}.
Finally, Eqs.\ \eqref{eq:TLCRL} follow upon application of 
$\widehat\nabla_{\widehat A\widehat\ad}$
on the appropriate sections.

\hfill $\Box$

\paragraph{Levi-Civita connection.}\label{par:LCC}
In the class of affine connections on $\CM$ which have totally trace-free
torsion tensors $T^\mp$ there exists a supersymmetric analog of
the {\it Levi-Civita connection}. 

{\Pro\label{thm:LCC}
Let $\CM$ be an RC supermanifold. For a given pair of non-degenerate 
sections $e\in H^0(\CM,\Lambda^2\CH^\vee)$ and 
$\tilde\varepsilon\in H^0(\CM,\Lambda^2\widetilde\CS^\vee)$
there always exists 
a unique torsion-free connection
$D$ on $\CM$  such that
$$ D e\ =\ 0\ =\ D\tilde\varepsilon.$$
In addition,
there is a unique scale (up to multiplicative constants) for which
this connection coincides with the one given by Prop.\ \ref{pro:LC1} 
} \vspace*{10pt}

\noindent
{\it Proof:}
One first notices that 
$$g\ :=\ e\otimes\tilde\varepsilon\in 
H^0(\CM,\Lambda^2\CH^\vee\otimes\Lambda^2\widetilde\CS^\vee)\ \subset\ 
H^0(\CM,\Omega^1\CM\odot \Omega^1\CM)$$ 
can be regarded as a holomorphic
metric on $\CM$ which, in fact, reduces to an ordinary holomorphic metric on 
$\CM_{\rm red}$.\footnote{Since  
$e$ and $\tilde\varepsilon$ are assumed to be non-degenerate, their corresponding
matrix representations are of full rank and hence as matrices they are invertible.}
Since $D(e\otimes\tilde\varepsilon)=(De)\otimes\tilde\varepsilon+
e\otimes(D\tilde\varepsilon)$, we have further that
$D g=0$. Together with the condition of vanishing torsion,
the proof reduces to that one familiar from ordinary Riemannian geometry (modulo
changes of signs due to the $\IZ_2$-grading).

Next one realizes that
$$
{\rm Ber}\,\CH\ \cong\ ({\rm Ber}\,\Lambda^2\CH)^{1/(1-\CN)},
$$
which
can be deduced from the definition of the Berezinian sheaf by using splitting principle 
arguments,
for instance. Hence,
$$
 {\rm Ber}\,\CH^\vee\ \cong\ ({\rm Ber}\,\Lambda^2\CH^\vee)^{1/(1-\CN)}\ 
      \cong\ {\rm Ber}\,\widetilde\CS^\vee.
$$
Thus, there exists 
a unique scale (up to multiplicative constants) where $D$ annihilates both,
$e\in H^0(\CM,\Lambda^2\CH^\vee)$ and $\varepsilon\in H^0(\CM,{\rm Ber}\,\CH^\vee)$.
Hence, by the uniqueness (for $\CN\neq4$) shown in Prop.\ \ref{pro:LC1}, $D$ coincides with $\nabla$.

\hfill $\Box$ 

\noindent
Hence, $\CM$ equipped with that type of connection
is a complex quaternionic RC supermanifold.
In full analogy with ordinary Riemannian geometry, we shall refer to this 
connection as the Levi-Civita connection. 

\paragraph{$\CN=4$ case.}\label{par:N4C}
As shown in Prop.\ \ref{pro:LC1}, there is no {\it unique} connection
$\nabla$ for $\CN=4$ which is solely
determined by the requirements of having
totally trace-free torsion and simultaneously 
annihilating both volume forms on $\CH$ and $\widetilde\CS$. To jump ahead of 
our story a bit, working with such a connection would result in 
a dependence of the supertwistor space $\CP$ associated with an RC
complex quaternionic supermanifold $\CM$
on the chosen scale on the latter. Of course, the definition of 
$\CP$ should only depend on the (super)conformal class of $\CM$, that is, it should
be independent of the particular scale. 

Nevertheless, as seen above, 
the Levi-Civita connection $D$ will always exist no matter what the chosen
value of $\CN$ is. Moreover, if $\CN=4$, it is possible to compute the change
of the Levi-Civita connection under superconformal rescalings since the 
usual torsion obstructions disappear. 

{\Pro\label{pro:CoSLC}
 Let $\CM$ be a $(4|8)$-dimensional RC supermanifold equipped with the
Levi-Civita connection. Under a rescaling of the form
$\tilde\varepsilon\mapsto\g\tilde\varepsilon$, where $\g$ is a nonvanishing
holomorphic function, the change of the Levi-Civita connection $D$
to the new one $\widehat D$ is given by the following contorsion tensors:
\begin{equation}\label{eq:changecon-LC}
{\Theta_{A\ad B}}^C\ =\ -\g_{A\ad}{\d_B}^C\qquad{\rm and}\qquad
{\Theta_{A\ad\bd}}^\gd\ =\ \gamma_{A\ad}{\d_\bd}^\gd
\end{equation}
Here, $\g_{A\ad}:=E_{A\ad}\log\g$, as before. This implies that the
new connection $\widehat D_{\widehat A\widehat\ad}=E_{\widehat A\widehat\ad}\lrcorner\widehat D
=E_{A\ad}\lrcorner\widehat D$
acts as follows:
\begin{equation}\label{eq:TLCRL-LC}
\begin{aligned}
  D_{A\ad}\mu^B\ &\mapsto\ \widehat D_{\widehat A\widehat\ad}\mu^{\widehat B}\kern-7pt &&=\ 
  D_{A\ad}\mu^B-\g_{A\ad}\mu^B,\\
  D_{A\ad}\lambda^\bd\ &\mapsto\ 
   \widehat D_{\widehat A\widehat\ad}\lambda^{\widehat\bd}\kern-7pt &&=\ 
  D_{A\ad}\lambda^\bd+\g_{A\ad}\lambda^\bd,\\
  D_{A\ad}\mu_B\ &\mapsto\ 
      \widehat D_{\widehat A\widehat\ad}\mu_{\widehat B}\kern-7pt &&=\
  D_{A\ad}\mu_B+\gamma_{A\ad}\mu_B,\\ 
  D_{A\ad}\lambda_\bd\ &\mapsto\ 
     \widehat D_{\widehat A\widehat\ad}\lambda_{\widehat\bd}\kern-7pt &&=\
  D_{A\ad}\lambda_\bd-\g_{A\ad}\lambda_\bd,\\ 
\end{aligned}
\end{equation}
where $\mu^A$ and $\lambda^\ad$ 
are sections of the vector bundles $\CH$ and $\widetilde\CS$,
respectively, together with $\mu^{\widehat A}=\mu^A$
and $\lambda^{\widehat\ad}=\lambda^\ad$ and similarly for their duals.
  }\vspace*{10pt}

\noindent
{\it Proof:} Under a change of scale $\tilde\varepsilon\mapsto
\widehat{\tilde\varepsilon}=\g\tilde\varepsilon$, the symplectic two-form
$e\in H^0(\CM,{\rm Ber}\,\Lambda^2\CH^\vee)$ behaves 
as $$e\ \mapsto\ \widehat e\ =\ \g^{\frac{2}{2-\CN}}e\ =\ \g^{-1}e.$$ 
This simply follows from the isomorphisms
${\rm Ber}\,\CH^\vee\cong({\rm Ber}\,\Lambda^2\CH^\vee)^{1/(1-\CN)} 
  \cong{\rm Ber}\,\widetilde\CS^\vee$. 
Hence, $g=e\otimes\tilde\varepsilon\mapsto\widehat g=g$. Note that $\kappa$ 
(see Eq.\ \eqref{eq:defkap}) vanishes identically for $\CN=4$. In addition, we have
$$ \widehat D\widehat e\ =\ 0\ =\ \widehat D\widehat{\tilde\varepsilon}. $$
Hence, the induced contorsion tensor ${\Theta_{A\ad B\bd}}^{C\gd}$ must be zero,
i.e. $\widehat D = D$ upon action on $T\CM$ (this result was already expected
by Eqs.\ \eqref{eq:beriso1} and \eqref{eq:beriso} for $\CN=4$).
It is then rather 
straightforward to verify that the above conditions imply Eqs.\ \eqref{eq:changecon-LC}.

\hfill $\Box$

\paragraph{Curvature.}
Given any connection $\nabla$ on $\CM$, the associated curvature 
two-form 
\begin{equation}
R\ =\ ({R_\ua}^\ub)\ =\ 
(\tfrac{1}{2}E^\ud\wedge E^\uc {R_{\uc\ud\ua}}^\ub),
\end{equation}
which takes
values in $\mathfrak{g}$, is defined by
\begin{equation}
 {R_\ua}^\ub\ =\ \dt{\Omega_\ua}^\ub+{\Omega_\ua}^\uc\wedge{\Omega_\uc}^\ub.
\end{equation}
The components of the curvature read explicitly as
\begin{equation}
\begin{aligned}
 {R_{\ua\ub\uc}}^\ud\ =\ &E_\ua{\Omega_{\ub\uc}}^\ud-(-)^{p_\ua p_\ub}
   E_\ub{\Omega_{\ua\uc}}^\ud+(-)^{p_\ua(p_\ub+p_\uc+p_\ue)}
   {\Omega_{\ub\uc}}^\ue{\Omega_{\ua\ue}}^\ud\ -\\
     &\kern1cm-(-)^{p_\ub(p_\uc+p_\ue)}{\Omega_{\ua\uc}}^\ue{\Omega_{\ub\ue}}^\ud-
   {f_{\ua\ub}}^\ue{\Omega_{\ue\uc}}^\ud. 
\end{aligned}
\end{equation}
In addition, torsion and curvature are combined into the standard formula
\begin{equation}
  [\nabla_\ua,\nabla_\ub\}u^\ud\ =\ (-)^{p_\uc(p_\ua+p_\ub)} u^\uc 
 {R_{\ua\ub\uc}}^\ud-{T_{\ua\ub}}^\uc\nabla_\uc u^\ud.
\end{equation}
Here, $u^\ua$ is some tangent vector on $\CM$. This equation might concisely be 
rewritten as
\begin{equation}\label{eq:curv}
  [\nabla_\ua,\nabla_\ub\}\ =\ R_{\ua\ub}-{T_{\ua\ub}}^\uc\nabla_\uc ,
\end{equation}
where $R_{\ua\ub}={R_{\ua\ub C}}^D {M_D}^C+{R_{\ua\ub\gd}}^\dd {M_\dd}^\gd$ 
together with the generators ${M_A}^B$ and ${M_\ad}^\bd$ of the Lie superalgebra
$\mathfrak{g}$. 

Note that because of the factorization $T\CM\cong\CH\otimes\widetilde\CS$, we have
\begin{equation}
R\ =\ R_\CH\otimes{\rm id}_{\widetilde\CS}+{\rm id}_\CH\otimes R_{\widetilde\CS}.
\end{equation}
Here, $R_\CH$ can be viewed as a section of $\Lambda^2\Omega^1\CM\otimes{\rm End}\,\CH$
while $R_{\widetilde\CS}$ as a section of 
$\Lambda^2\Omega^1\CM\otimes{\rm End}\,\widetilde\CS$.
In the structure frame, the decomposition of $R$ looks as
\begin{equation}\label{eq:curvdec}
{R_\ua}^\ub={R_{A\ad}}^{B\bd}\ =\ 
 {R_A}^B{\d_\ad}^\bd+{\d_A}^B{R_\ad}^\bd.
\end{equation}
Furthermore, recalling Eq.\ \eqref{eq:2formdec}, we have further decompositions
of $R$ (respectively, of $R_\CH$ and $R_{\widetilde\CS}$) into
$R^\mp$ (respectively, into $R^\mp_\CH$ and $R^\mp_{\widetilde\CS}$).

{\Pro\label{pro:curv}
 Let $\CM$ be a complex quaternionic
 RC supermanifold.  In the structure frame,
the curvature parts $R^\mp_\CH$ and $R^\mp_{\widetilde\CS}$ of 
$R^\mp$ are of the following form:
\begin{equation}\label{eq:procurv}
\begin{aligned}
 R^-_\CH\ &:\ {R_{A(\ad B\bd)C}}^D\kern-7pt &&=\
  -2(-)^{p_C(p_A+p_B)}R_{C[A\ad\bd}{\d_{B\}}}^D,\\
 R^+_\CH\ &:\ \epsilon_{\ad\bd}{R_{ABC}}^D\kern-7pt &&=\ \epsilon_{\ad\bd}(
 {C_{ABC}}^D-2(-)^{p_C(p_A+p_B)}\Lambda_{C\{A}{\d_{B]}}^D),\\
 R^-_{\widetilde\CS}\ &:\ {R_{A(\ad B\bd)\gd}}^\dd\kern-7pt
&&=\ {C_{AB\ad\bd\gd}}^\dd+2\Lambda_{AB}{\d_{(\ad}}^\dd\epsilon_{\bd)\gd},\\
 R^+_{\widetilde\CS}\ &:\ \epsilon_{\ad\bd}{R_{AB\gd}}^\dd,\kern-7pt&&
\end{aligned}
\end{equation}
where $R_{AB\ad\bd}:={R_{AB\ad}}^\gd\epsilon_{\gd\bd}$ and
$$
\begin{aligned}
 {C_{ABC}}^D\ =\ {C_{\{ABC]}}^D,\quad
 (-)^C {C_{ABC}}^C\ =\ 0,\quad\Lambda_{AB}\ =\ \Lambda_{[AB\}},\quad
 R_{AB\ad\bd}\ =\ R_{\{AB](\ad\bd)},\\
{C_{AB\ad\bd\gd}}^\dd\ =\ {C_{[AB\}(\ad\bd\gd)}}^\dd,\quad
 {C_{AB\ad\bd\gd}}^\gd\ =\ 0.\kern3cm
\end{aligned}
$$
In addition, the Ricci tensor ${\rm Ric}_{A\ad B\bd}:=(-)^{p_C+p_Cp_B} 
{R_{A\ad C\gd B\bd}}^{C\gd}$ is
given by
\begin{equation}\label{eq:riccitensor}
 {\rm Ric}_{A\ad B\bd}\ =\ -(2-\CN)R_{AB\ad\bd}+(6-\CN)\Lambda_{AB}\epsilon_{\ad\bd},
\end{equation}
where ${\rm Ric}_{A\ad B\bd}=(-)^{p_Ap_B}{\rm Ric}_{B\bd A\ad}$.
}\vspace*{10pt}

\noindent
{\it Proof:}
The proof is based on Bianchi identities and certain index symmetries of the 
curvature
tensor. However, the calculations are rather technical and lengthy, and therefore 
postponed to 
App.\ \ref{app:curv}.

\hfill $\Box$

\noindent
In the following, we shall refer to the quantity $\Lambda_{AB}$ as the
{\it cosmological constant}.

\subsection{\kern-10pt Self-dual supergravity equations}\label{sec:sdsg}

\paragraph{Self-duality.}
Let $\CM$ be a complex quaternionic RC supermanifold which is
equipped with the Levi-Civita connection. 
It is called {\it self-dual Einstein} if ${C_{AB\ad\bd\gd}}^\dd=0$ and
simultaneously 
${R_{AB\ad}}^\bd=0$. 

{\Def\label{def:CQKRC}
 A complex quaternionic RC supermanifold is said to be complex
quaternionic K\"ahler if it is equipped with
the Levi-Civita connection and is also 
self-dual Einstein.}\vspace*{10pt}

\noindent
If, in addition, the
cosmological constant $\Lambda_{AB}$ vanishes as well, we call $\CM$ {\it
self-dual}. In the latter case, the curvature $R$ is of the
form $R=R^+_\CH\otimes{\rm id}_{\widetilde\CS}$, i.e.
\begin{equation}\label{eq:sdsg}
[D_{A\ad},D_{B\bd}\}\ =\ \epsilon_{\ad\bd}R_{AB},
\end{equation}
where $R_{AB}$ is of the form
$R_{AB}={C_{ABC}}^D{M_D}^C$.
Furthermore, the connection $D$ has components
\begin{equation}
 D_{A\ad}\ =\ {E_{A\ad}}^{M\bd}\partial_{M\bd}+{\Omega_{A\ad B}}^C{M_C}^B.
\end{equation}
Obviously, this says that $D$ on $\widetilde\CS$ of 
$T\CM\cong\CH\otimes\widetilde\CS$ is flat.
 It should be
noticed that the superfield components of $R_{AB}$ are not independent of
each other because of the Bianchi identities
\begin{equation}\label{eq:BI}
 D_{[A\ad}R_{B\}C}\ =\ 0.
\end{equation}
The field equations of self-dual supergravity with vanishing cosmological
constant then follow from these identities together with
\eqref{eq:sdsg}. Their explicit form can be found in Siegel \cite{Siegel:1992wd}.
We may summarize by giving the following definition:

{\Def\label{def:CHKRC}
 A complex quaternionic K\"ahler RC supermanifold is called a complex 
hyper-K\"ahler RC supermanifold 
 if the Levi-Civita connection on $\widetilde\CS$ is flat.}\vspace*{10pt}

\noindent
In addition, Prop.\ \ref{pro:curv} shows that if 
$\CM$ is self-dual, it is also Ricci-flat. Altogether, a complex 
hyper-K\"ahler RC supermanifold $\CM$ is Ricci-flat and has trivial Berezinian
sheaf ${\rm Ber}(\CM)$, i.e. it is a
{\it Calabi-Yau} supermanifold. In this respect, it is worth
mentioning that contrary to ordinary
complex manifolds, complex supermanifolds with trivial Berezinian sheaf do 
not automatically admit Ricci-flat metrics (see e.g. 
Refs.\ \cite{Rocek:2004bi}). We shall refer to this latter type of supermanifolds
as {\it formal} Calabi-Yau supermanifolds. Furthermore, for an earlier
account  
of hyper-K\"ahler supermanifolds of dimension $(4k|2k+2)$,
though in a slightly different setting, 
see Merkulov \cite{Merkulov:1992}. See also Lindstr\"om et al. 
\cite{Lindstrom:2005uh}.

\paragraph{Second Plebanski equation.}
By analyzing the constraint equations
\eqref{eq:sdsg} in a noncovariant gauge called {\it light-cone gauge},
Siegel \cite{Siegel:1992wd}
achieved reducing them to a single equation on a superfield
$\Theta$, which in fact
 is the supersymmetrized analog of Plebanski's second equation 
\cite{Plebanski:1975wn}.
In particular, in this gauge
the vielbeins turn out to be 
\begin{equation}\label{eq:siegelvielbein}
\begin{aligned}
 {E_{A\dot1}}^{M\bd}\ &=\ {\d_A}^M{\d_{\dot1}}^\bd+
  \tfrac{1}{2}(-)^{p_D}{\d_A}^N{\d_B}^O
  (\partial_{N\dot2}\partial_{O\dot2}\Theta)\omega^{BC}{\d_C}^M{\d_{\dot2}}^\bd,\\
  {E_{A\dot2}}^{M\bd}\ &=\ {\d_A}^M{\d_{\dot2}}^\bd,
\end{aligned}
\end{equation}
where $(\partial_{M\ad})=(\partial_{\mu\ad},\partial_{m\ad})$ with
$\partial_{\mu\ad}:=\partial/\partial x^{\mu\ad}$ and
$\partial_{m\ad}:=\partial/\partial\eta^{m\ad}$ and $\omega^{AB}:=(\epsilon^{\a\b},\d^{ij})$. 
By a slight abuse of
notation, we shall write $\partial_{A\ad}\equiv{\d_A}^M\partial_{M\ad}$ in
the following. Furthermore, the components of the connection one-form
in this gauge are given by 
\begin{equation}\label{eq:siegelconnection}
\begin{aligned}
 {\Omega_{A\dot1 B}}^C\ &=\ -(-)^{p_D}\tfrac{1}{2}
  \partial_{A\dot2}\partial_{B\dot2}\partial_{D\dot2}\Theta\, 
  \omega^{DC},\\
 {\Omega_{A\dot2 B}}^C\ &=\ 0.
\end{aligned}
\end{equation}
The equation $\Theta$ is being subject to is then 
\begin{equation}\label{eq:superplebanski}
\epsilon^{\ad\bd}\partial_{A\ad}\partial_{B\bd}\Theta+\tfrac{1}{2}(-)^{p_C}
 (\partial_{A\dot2}\partial_{C\dot2}\Theta)\omega^{CD}
 (\partial_{D\dot2}\partial_{B\dot2}\Theta)\ =\ 0.
\end{equation}
In summary, the field equations of self-dual supergravity
in light-cone gauge are equivalent to \eqref{eq:superplebanski}. 

\paragraph{Another formulation.}
Subject of this paragraph is to give another (equivalent) formulation
of the self-dual supergravity equations with vanishing cosmological
constant. 
The following proposition
generalizes results of Mason and Newman \cite{Mason:1989ye} to the
supersymmetric situation.

{\Pro
Let $\CM$ be a complex quaternionic RC supermanifold which is equipped with the
Levi-Civita connection. Suppose further we are given
vector fields $V_{A\ad}$ on $\CM$ which obey
\begin{equation}\label{eq:vectsdsg}
 [V_{A(\ad},V_{B\bd)}\}\ =\ 0.
\end{equation}
Then \eqref{eq:vectsdsg} is an equivalent formulation of
the self-dual supergravity equations, i.e. given vector fields $V_{A\ad}$ on $\CM$ satisfying 
\eqref{eq:vectsdsg}, it is always possible
to find frame fields $E_\ua$ such that the self-dual supergravity equations 
with zero cosmological constant are satisfied thus
making $\CM$ into a complex hyper-K\"ahler RC supermanifold. Conversely, given
a complex hyper-K\"ahler RC supermanifold $\CM$, then there will always
exist vector fields $V_{A\ad}$ on $\CM$ which satisfy \eqref{eq:vectsdsg}.}\vspace*{10pt}

\noindent{\it Proof:}
In fact, it is not too difficult to see that \eqref{eq:vectsdsg} 
implies the self-dual supergravity equations with $\Lambda_{AB}=0$. 
Indeed, by Frobenius' theorem (see e.g. Manin \cite{Manin} for the 
case of supermanifolds) we may choose coordinates such that the $V_{A\dot2}$s become
coordinate derivatives, i.e.
$$ V_{A\dot2}\ =\ \partial_{A\dot2}. $$ 
In addition, by choosing a gauge such that the $V_{A\dot1}$s take the form
$$ V_{A\dot1}\ =\ \partial_{A\dot1}+\tfrac{1}{2}(-)^{p_B}
  (\partial_{A\dot2}\partial_{B\dot2}\Theta)
   \omega^{BC}\partial_{C\dot2},
$$
where $\Theta$ is some to be determined superfield, all equations \eqref{eq:vectsdsg}  but one
are identically satisfied. In particular, only $[V_{A\dot1},V_{B\dot1}\}=0$
gives a nontrivial condition on $\Theta$. In fact, this equation reduces to 
\eqref{eq:superplebanski}. Therefore, taking the vielbeins and the components
of the connection one-form as  in \eqref{eq:siegelvielbein} and \eqref{eq:siegelconnection},
respectively, we arrive at the desired result.

Conversely, given some complex hyper-K\"ahler RC supermanifold $\CM$, the only nonvanishing
components of the connection
one-form are ${\Omega_{A\ad B}}^C$. By virtue of the vanishing of the torsion, 
Eqs.\ \eqref{eq:torsioncomp} imply
$$
 {f_{\ua\ub}}^\uc\ =\  {\Omega_{\ua\ub}}^\uc-(-)^{p_\ua p_\ub}{\Omega_{\ub\ua}}^\uc.
$$
Since ${\Omega_{\ua\ub}}^\uc={\Omega_{A\ad B\bd}}^{C\gd}
={\d_\bd}^\gd{\Omega_{A\ad B}}^C$, we find
$$
 {f_{A(\ad B\bd)}}^{C\gd}\ =\ {\d_{(\ad}}^\gd{\Omega_{[A\bd)B\}}}^C.
$$
By the discussion
given in the last paragraph of Sec.~\ref{sec:sdsg}, we know that there exists a gauge
in which ${\Omega_{[A\ad B\}}}^C$ vanishes. Therefore, there will always
exist vector fields $V_{A\ad}$ obeying \eqref{eq:vectsdsg}. 
This concludes the proof.

\hfill $\Box$ 

\section{\kern-10pt Twistor theory}\label{sec:TT}

Above we have introduced and discussed complex
quaternionic K\"ahler and hyper-K\"ahler RC supermanifolds by
starting from complex quaternionic
RC supermanifolds. In
this section, we shall be concerned with their twistorial description.
We first construct the supertwistor space, denoted by $\CP$, of a 
complex quaternionic RC supermanifold $\CM$. However,
as in the purely bosonic situation, we shall see that this will only work 
if one makes certain additional assumptions about the properties
of $\CM$. Having presented 
this construction, we then
show which additional structures on $\CP$ are needed to render
$\CM$ into
a complex hyper-K\"ahler RC supermanifold. We further give
an alternative formulation and eventually conclude this section
by introducing the bundle of local supertwistors.

\subsection{\kern-10pt Supertwistor space $(\CN\neq4)$}\label{sec:STC}

\paragraph{Conic structures.}
In order to proceed in finding an appropriate twistor 
description, so-called {\it conic structures} 
appear to be an adequate tool. Let us therefore recall their definition.

{\Def [Manin \cite{Manin}] Let $\CM$ be a complex supermanifold with
holomorphic tangent bundle $T\CM$. A $(p|q)$-conic structure
on $\CM$ is a closed subsupermanifold $\CF$ in the relative Gra{\ss}manian
$G_{\CM}(p|q; T\CM)$, 
$$
G_{\CM}(p|q; T\CM)\ :=\ \{\text{rank } p|q\text{ local direct summands of } T\CM\},
$$
such that the projection $\pi:\CF\to\CM$ is a
submersion.
}\vspace*{10pt}

\noindent
Putting it differently, at any point $x\in\CM$ such an $\CF$ determines
a set of $(p|q)$-dimensional tangent spaces in the fibre of $T\CM$ over
$x$ corresponding to the points $\pi^{-1}(x)\subset 
G_{\CM}(p|q; T_x\CM)$.

\paragraph{$\beta$-plane bundle.}\label{par:bpb}
Having given this definition, we may now introduce a canonical conic
structure on a complex quaternionic RC supermanifold $\CM$. Recall again from
\eqref{eq:ES1} that the tangent bundle $T\CM$ 
of $\CM$ is of the form
$$
 0\ \longrightarrow\ \CE\otimes\widetilde\CS\ \longrightarrow\  
   T\CM\ \longrightarrow\ \CS\otimes\widetilde\CS\ \longrightarrow\ 0,
$$
where $\CS$ and $\widetilde\CS$ are both of rank $2|0$ and $\CE$ is of rank
$0|\CN$. 

Let now $\CF$ be the relative projective line bundle 
$P_\CM(\widetilde\CS^\vee)$ on $\CM$. Then the above sequence 
induces a canonical $(2|\CN)$-conic structure on $\CM$, that
is, an embedding $\CF\hookrightarrow G_\CM(2|\CN; T\CM)$. In local coordinates, it is given by
\begin{equation}\label{eq:cancon}
\begin{aligned}
 \CF\ &\to\ G_\CM(2|\CN; T\CM),\\
 [\lambda_\ad]\ &\mapsto\ \Scr{D}\ :=\ \langle \lambda^\ad E_{A\ad}\rangle.
\end{aligned}
\end{equation}
Here, $[\lambda_\ad]$ are homogeneous fibre coordinates of 
$\pi_1:\CF\to\CM$ and the $E_{A\ad}$s are the frame fields on $\CM$. This
construction leads naturally to the following definition:

{\Def A $\beta$-surface $\Sigma$ in a complex 
quaternionic RC supermanifold $\CM$ is 
a complex subsupermanifold of dimension $(2|\CN)$ with the property that 
at each point $x\in\Sigma$, the tangent space $T_x\Sigma$ is spanned by vectors of the form
\eqref{eq:cancon}, where $\lambda_\ad$ is fixed up to rescalings.} \vspace*{10pt}

\noindent
This in particular means that the components of a tangent vector on
$\Sigma$ are always of the form $\mu^A\lambda^\ad$, where $\mu^A$ is arbitrary.

It is worth noting that on $\CM_{\rm red}$ this notion of $\beta$-surfaces
reduces to the standard one (see e.g. Refs.\ \cite{Ward:1990vs,Mason:1991rf}). 
Next we introduce the notion of {\it right-flatness}.

{\Def A complex quaternionic RC supermanifold $\CM$ is said to be right-flat, 
if the $R_{A(\ad B\bd\,\gd\dd)}$-components of the curvature tensor 
vanish.} \vspace*{10pt}

\noindent
Clearly, for $\CN=0$ this reduces to the
standard definition of the vanishing of the anti-self-dual part of the
Weyl tensor.
Now we are in the position to prove the following proposition:

{\Pro\label{pro:PRF} Let $\CM$ be a complex quaternionic RC
supermanifold.
For any given point of $\CF=P_\CM(\widetilde\CS^\vee)$ there exists a corresponding
$\beta$-surface $\Sigma$ in $\CM$ if and only if 
$\CM$ is right-flat.}\vspace*{10pt}

\noindent
Putting it differently, the distribution defined by \eqref{eq:cancon} is
integrable, i.e. closed under the graded Lie bracket, if and only if $\CM$ is 
right-flat.\vspace*{10pt}

\noindent
{\it Proof:} 
It is not too difficult to show that the graded Lie bracket 
of two vector fields $E_A:=\lambda^\ad E_{A\ad}$ and $E_B:=\lambda^\bd E_{B\bd}$ is given 
by
$$
 [E_A,E_B\}
 \ =\ 2\big(\lambda^\ad(\nabla_{[A\ad}\lambda^\bd)E_{B\}\bd}-\lambda^\ad\lambda^\bd
   {T_{A(\ad B\bd)}}^{C\gd}E_{C\gd}-\lambda^\ad{\Omega_{[A\ad B\}}}^C E_C\big).
$$
However, by virtue of the vanishing of the torsion, the second term
on the right-hand side of this equation vanishes identically. Hence, the distribution
generated by $E_A$ is integrable if and only if
\begin{equation}\label{eq:lifteq}
 \lambda^\ad\lambda^\bd\nabla_{A\ad}\lambda_\bd\ =\ 0.
\end{equation}
Since the integrability condition of this equation is equivalent to the vanishing of
the curvature components  $R_{A(\ad B\bd\,\gd\dd)}$, we arrive at the desired
result. 

A remark is in order: if $\CM$ was not complex quaternionic but only equipped
with a connection whose torsion is totally trace-free (cf. our discussion given in 
\ref{par:CTC2}~\,\,),
then the above requirement of the integrability of the distribution $\Scr{D}$ would enforce
the vanishing of ${T_{A(\ad B\bd)}}^{C\gd}$ as the first term appearing in the equation for
$[E_A,E_B\}$ is proportional to the trace. Hence, the whole torsion tensor 
${T_{A\ad B\bd}}^{C\gd}$ would be zero
(see Eq.\ \eqref{eq:redtors}). By virtue of Eq.\ \eqref{eq:lifteq}, $\CM$ would then become
a  right-flat complex quaternionic RC supermanifold.

\hfill $\Box$ 

\noindent
Note that $\lambda_\ad$ can be normalized such that Eq.\ \eqref{eq:lifteq} becomes,
\begin{equation}\label{eq:lifteqre}
 \lambda^\ad\nabla_{A\ad}\lambda_\bd\ =\ 0,
\end{equation}
i.e. $\lambda_\ad$ is covariantly constant (i.e. $\lambda_\ad$ is an auto-parallel
co-tangent spinor)
on $\Sigma\hookrightarrow\CM$. In addition,
we point out that this equation is scale invariant.
This follows from the transformation laws \eqref{eq:TLCRL} of
the connection under rescalings. For $\CN=4$, this equation is not scale
invariant as $\nabla$ is not unique. Therefore, its solutions  
may depend on the chosen scale on $\CM$. We shall address this issue in more
detail later on.

Following the terminology of Mason and Woodhouse \cite{Mason:1991rf}, we shall call $\CF$ the
{\it $\beta$-plane bundle}. We also refer to $\CF$ as the {\it correspondence space}. 

\paragraph{Supertwistor space.}
Note that $\beta$-surfaces $\Sigma$ lift into $\CF$ and in addition also foliate $\CF$. 
The lift $\widetilde\Sigma$ of $\Sigma$ is a section of $\CF|_\Sigma\to\Sigma$ satisfying
Eqs.\ \eqref{eq:lifteq}. The tangent vector fields on $\widetilde\Sigma$ are then given by
\begin{equation}
 \widetilde{E}_A\ =\ E_A+\lambda^\ad\lambda_\gd{\Omega_{A\ad\bd}}^\gd\der{\lambda_\bd}.
\end{equation}
Therefore, we canonically obtain an integrable rank-$2|\CN$ distribution $\Scr{D}_\CF\subset T\CF$
on the correspondence space generated by the $\widetilde E_A$s, i.e.
 $\Scr{D}_\CF=\langle\widetilde E_A\rangle$.
We shall refer to $\Scr{D}_\CF$ as the {\it twistor distribution}.
After quotienting $\CF$ by the twistor distribution,
we end up with the following double fibration:
\begin{equation}\label{eq:DF2}
\begin{aligned}
\begin{picture}(50,40)
\put(0.0,0.0){\makebox(0,0)[c]{$\CP$}}
\put(64.0,0.0){\makebox(0,0)[c]{$\CM$}}
\put(34.0,33.0){\makebox(0,0)[c]{$\CF$}}
\put(7.0,18.0){\makebox(0,0)[c]{$\pi_2$}}
\put(55.0,18.0){\makebox(0,0)[c]{$\pi_1$}}
\put(25.0,25.0){\vector(-1,-1){18}}
\put(37.0,25.0){\vector(1,-1){18}}
\end{picture}
\end{aligned}
\end{equation}
Here, $\CP$ is a $(3|\CN)$-dimensional complex supermanifold which we call the {\it supertwistor
space} of $\CM$. Note that this construction is well-defined if we additionally 
assume that $\CM$ is {\it civilized}, that is, 
 $\CP$ is assumed to have the same topology as the supertwistor 
space associated with any convex region in flat
superspace $\IC^{4|2\CN}$. Otherwise, one my end up with non-Hausdorff
spaces;
see e.g. Ward and Wells \cite{Ward:1990vs} and Mason and Woodhouse \cite{Mason:1991rf} 
for a discussion in the purely
bosonic situation. 
Moreover, without this convexity assumption, the Penrose transform, which relates
certain cohomology groups on $\CP$ to solutions to certain partial differential
equations on $\CM$, will not be an isomorphism (see also Sec.\ \ref{subsec:EF}).

By virtue of this double fibration, we have a geometric correspondence between the two
supermanifolds $\CM$ and $\CP$. In particular, any point $x\in\CM$ is associated with the
set $\pi_2(\pi_1^{-1}(x))$ in $\CP$ consisting of all $\beta$-surfaces being {\it incident} 
with $x$. Conversely, any point $z$ in supertwistor space $\CP$ corresponds to an
$\beta$-surface $\pi_1(\pi_2^{-1}(z))$ in $\CM$. As $\CF\to\CM$
is a $\IP^1$-bundle over $\CM$, the submanifolds $\pi_2(\pi_1^{-1}(x))$ are biholomorphically
equivalent to $\IP^1$ and are parametrized by $x\in\CM$.

We may now state the following basic result:

{\Thm\label{thm:T1}
There is a one-to-one correspondence between:
\begin{itemize}
\setlength{\itemsep}{-1mm}
\item[(i)] civilized right-flat complex quaternionic RC supermanifolds $\CM$ of dimension
           $(4|2\CN)$ and
\item[(ii)] $(3|\CN)$-dimensional complex supermanifolds $\CP$ each containing a 
            family of rational cur\-ves
            biholomorphically equivalent to $\IP^1$ and with normal
            bundle $\cN_{\IP^1|\CP}$ inside $\CP$  described by 
            \begin{equation}\label{eq:normalbundle}
            0\ \longrightarrow\ \Pi\CO_{\IP^1}(1)\otimes\IC^\CN\ \longrightarrow\  
                  \cN_{\IP^1|\CP}\ \longrightarrow\ \CO_{\IP^1}(1)\otimes\IC^2\ \longrightarrow\ 0,
            \end{equation}
            where $\CO_{\IP^1}(1)$ is
            the dual tautological $(c_1=1)$ bundle on $\IP^1$ and $\Pi$ is the Gra{\ss}mann parity 
            changing
            functor.
\end{itemize}
}\vspace*{10pt}

\noindent
{\it Proof:} 
Let us first show (i) $\to$ (ii): In fact, we have already seen that for any 
complex quaternionic RC
supermanifold $\CM$ with the above properties, 
   there always exists an associated $(3|\CN)$-dimensional complex 
supermanifold $\CP$ containing holomorphically embedded projective lines
$\pi_2(\pi_1^{-1}(x))\cong\IP^1$ for $x\in\CM$. It remains to verify that
each of it has a normal bundle $\cN_{\IP^1|\CP}$ of the above type. To show this, we notice that
 $\cN_{\IP^1|\CP}$ is described by the exact sequence
\begin{equation}\label{eq:aseq2}
 0\ \longrightarrow\ \Scr{D}_\CF\ \longrightarrow\  
   \pi_1^*T\CM\ \longrightarrow\ \pi_2^*\cN_{\IP^1|\CP}\ \longrightarrow\ 0,
\end{equation}
where $\Scr{D}_\CF$ is the twistor distribution. Clearly, the distribution
 $\Scr{D}_\CF$ is described by  
$$0\ \longrightarrow\ \CO_{\IP^1}(-1)\otimes\IC^2\ \longrightarrow\ \Scr{D}_\CF\ \longrightarrow \  
 \Pi\CO_{\IP^1}(-1)\otimes\IC^\CN\ \longrightarrow\ 0$$
when restricted to the fibres $\pi_1^{-1}(x)$ of $\CF\to\CM$. 
Furthermore, $\pi_1^*T\CM$ is trivial when restricted to 
$\pi_1^{-1}(x)$. Therefore, the maps of the above sequence are explicitly given by
$$
\begin{aligned}
 0\ \longrightarrow\ &\Scr{D}_\CF\ \longrightarrow\  
   &&\kern-7pt\pi_1^*T\CM\ \longrightarrow\ \pi_2^*\cN_{\IP^1|\CP}\ \longrightarrow\ 0,\\
                        &\ \mu^A\kern8pt \mapsto\ &&\kern-7pt\ \mu^A\lambda^\ad,\\
                &     &&\kern-7pt\ \ u^{A\ad}\ \ \ \ \mapsto\  \  u^{A\ad}\lambda_\ad,
\end{aligned}
$$
which completes the proof of the direction  (i) $\to$ (ii).

To show the reverse direction  (ii) $\to$ (i), one simply applies a supersymmetric version
of Kodaira's theorem of deformation
theory (Waintrob \cite{Waintrob:1986}). 
First, one notices that the obstruction group $H^1(\IP^1,\cN_{\IP^1|\CP})$ vanishes
which follows from the sequence \eqref{eq:normalbundle} and its induced long exact cohomology sequence:
$$
\begin{aligned}
    0\ &\longrightarrow\ H^0(\IP^1,\Pi\CO_{\IP^1}(1)\otimes\IC^\CN)\ \longrightarrow\
  H^0(\IP^1,\cN_{\IP^1|\CP})\ \longrightarrow\ \\ 
          &\kern10pt\longrightarrow\ H^0(\IP^1,\CO_{\IP^1}(1)\otimes\IC^2)\ \longrightarrow\  
         H^1(\IP^1,\Pi\CO_{\IP^1}(1)\otimes\IC^\CN)\ \longrightarrow\ \\
          &\kern20pt\longrightarrow\ H^1(\IP^1,\cN_{\IP^1|\CP})\ \longrightarrow\ 
           H^1(\IP^1,\CO_{\IP^1}(1)\otimes\IC^2)\ \longrightarrow\ 0.
 \end{aligned}
$$
Then there exists a dim$_\IC H^0(\IP^1,\cN_{\IP^1|\CP})=4|0+0|2\CN=4|2\CN$ parameter family $\CM$ of
deformations of $\IP^1$ inside $\CP$. 

If we let $\CF:=\{(z,\pi_2(\pi_1^{-1}(x)))\,|\,z\in\pi_2(\pi_1^{-1}(x)),\ z\in\CP,
\ x\in\CM\}\subset\CP\times\CM$, then
$\CF$ is a fibration over $\CM$. The typical fibres of $\CF\to\CM$ 
are complex projective lines $\IP^1$. Hence, we obtain a double fibration
$$
\begin{aligned}
\begin{picture}(50,40)
\put(0.0,0.0){\makebox(0,0)[c]{$\CP$}}
\put(64.0,0.0){\makebox(0,0)[c]{$\CM$}}
\put(34.0,33.0){\makebox(0,0)[c]{$\CF$}}
\put(7.0,18.0){\makebox(0,0)[c]{$\pi_2$}}
\put(55.0,18.0){\makebox(0,0)[c]{$\pi_1$}}
\put(25.0,25.0){\vector(-1,-1){18}}
\put(37.0,25.0){\vector(1,-1){18}}
\end{picture}
\end{aligned}
$$
where the fibres of $\CF\to\CP$ are $(2|\CN)$-dimensional complex 
subsupermanifolds of $\CM$.

Let $T\CF/\CP$ be the relative tangent sheaf on $\CF$ given by
$$
 0\ \longrightarrow\ T\CF/\CP\ \longrightarrow\
 T\CF\ \longrightarrow\ \pi_2^*T\CP\ \longrightarrow\ 0.
$$
Then (see above) we define a vector bundle $\cN$ on $\CF$ by
\begin{equation}\label{eq:aseqN}
 0\ \longrightarrow\ T\CF/\CP\ \longrightarrow\ \pi_1^*T\CM\ 
\longrightarrow\ \cN\ \longrightarrow\ 0.
\end{equation}
Clearly, the rank of $\cN$ is $2|\CN$ and furthermore, the restriction of $\cN$
to the fibre $\pi_1^{-1}(x)$ of $\CF\to\CM$ for $x\in\CM$ is isomorphic to the 
pull-back of the normal bundle of the curve $\pi_2(\pi_1^{-1}(x))\hookrightarrow\CP$.
Hence, $\cN$ may be identified with $\pi_2^*\cN_{\IP^1|\CP}$
and moreover, the relative tangent sheaf $T\CF/\CP$ with the twistor
distribution $\Scr{D}_\CF$. 

In addition, the bundle $\pi_1\,:\,\CF\to\CM$ is of the form $P_\CM(\widetilde\CS^\vee)$
for some rank $2|0$ vector bundle $\widetilde\CS$ (determined below) over $\CM$. Then we
denote by $\CO_\CF(-1)$ the tautological $(c_1=-1)$ bundle on $\CF$.  
It then follows from the above that 
the direct images\footnote{Given a mapping $\pi\,:\,\CM\to\cN$ of two
complex supermanifolds $\CM$ and $\cN$, the $q$-th direct image sheaf
$\pi^q_*\CE$ of a locally free sheaf $\CE$ over $\CM$ is defined by the
presheaf $\cN\supset\CU\ {\rm open}\ \mapsto H^q(\pi^{-1}(\CU),\CE)$
with the obvious restriction maps. The zeroth direct image sheaf
$\pi^0_*\CE$ is usually denoted by $\pi_*\CE$.}
 $\pi_{1*}(\Omega^1\CF/\CP\otimes\CO_\CF(-2))$ and
$\pi^1_{1*}(\Omega^1\CF/\CP\otimes\CO_\CF(-2))$ vanish. 
Therefore, we find that
$$ \pi_{1*}(T\CF/\CP)\ =\ 0\ =\ \pi^1_{1*}(T\CF/\CP)$$
upon application of Serre duality.\footnote{Recall that Serre duality asserts that for
any locally free sheaf $\CE$ on a compact complex manifold $M$ of dimension $d$, 
we have an isomorphism $H^q(M,\CE)\cong H^{d-q}(M,\Scr{K}_M\otimes\CE^\vee)$,
where $\Scr{K}_M$ is the canonical sheaf on $M$.}

Applying the direct image functor to the sequence \eqref{eq:aseqN}, we
obtain
$$
0\ \longrightarrow\ \pi_{1*}(T\CF/\CP)\ \longrightarrow\ \pi_{1*}(\pi_1^*T\CM)
 \ \longrightarrow\ \pi_{1*}\cN\ \longrightarrow\ \pi^1_{1*}(T\CF/\CP),
$$
and hence
$$
T\CM\ \cong\ \pi_{1*}\cN\ \cong\ \pi_{1*}(\pi_2^*\cN_{\IP^1|\CP}).
$$
Thus, the sequence \eqref{eq:normalbundle} yields
$$
\begin{aligned}
 0\ \longrightarrow\ \, &\pi_{1*}(\pi_2^*(\Pi\CO_{\IP^1}(1)\otimes\IC^\CN))\ \longrightarrow\  
  T\CM\ \longrightarrow\ \pi_{1*}(\pi_2^*(\CO_{\IP^1}(1)\otimes\IC^2))\ \longrightarrow\ 0\\
  &\kern2cm\parallel \kern3.18cm\parallel \kern3.25cm\parallel\\ 
 0\ &\kern.35cm\longrightarrow\ \kern.35cm\CE\otimes\widetilde\CS\kern.65cm\ 
  \longrightarrow\ \kern.65cm T\CM\kern.65cm\ \longrightarrow\kern.65cm\ \CS\otimes
  \widetilde\CS\kern.35cm\ \longrightarrow\kern.35cm\ 0
\end{aligned}
$$
since the first direct image sheaf $\pi^1_{*\,}(\pi_2^*(\Pi\CO_{\IP^1}(1)\otimes\IC^\CN))$ vanishes
(see above).
Above, we have introduced
$$\CS\ :=\ \pi_{1*}(\pi_2^*(\CO_{\IP^1}\otimes\IC^2),\quad
  \widetilde\CS\ :=\ \pi_{1*}(\pi_2^*(\CO_{\IP^1}(1))\quad{\rm and}\quad
  \CE\ :=\ \pi_{1*}(\pi_2^*(\Pi\CO_{\IP^1}\otimes\IC^\CN)
$$

Notice that by construction, the bundle $\CF\cong P_\CM(\widetilde\CS^\vee)$ is
an integrable $(2|\CN)$-conic structure on $\CM$.

\hfill $\Box$ 

\paragraph{Gindikin's two-forms and self-dual supergravity with $\Lambda_{AB}=0$.}
In this and the subsequent paragraph, we shall determine the structure on the
supertwistor space $\CP$ corresponding to a hyper-K\"ahler structure
on $\CM$. In view of that, recall that there always exists
a scale where the Levi-Civita connection $D$ coincides with
the connection $\nabla$.

Let $E^{A\ad}$ be the coframe fields on some complex quaternionic K\"ahler supermanifold
$\CM$. On the correspondence space
$\CF$ of the double fibration \eqref{eq:DF2}, we may introduce a differential
two-form
$\Sigma(\lambda)$ by setting
\begin{equation}\label{eq:gindikin}
  \Sigma(\lambda)\ :=\  E^{B\bd}\wedge E^{A\ad}\,e_{AB}
      \lambda_\ad\lambda_\bd,
\end{equation}
where $e\in H^0(\CM,\Lambda^2\CH^\vee)$ is assumed to be non-degenerate
and to obey $\nabla e=0$. 
Furthermore, let $\dt_h$ be the exterior
derivative on $\CF$ holding $\lambda_\ad$ constant. 

{\Pro 
There is a one-to-one correspondence between gauge equivalence classes
of solutions to the self-dual supergravity equations \eqref{eq:sdsg}
with vanishing cosmological constant on $\CM$ and 
equivalence classes of (global) $\dt_h$-closed non-degenerate 
differential two-forms $\Sigma(\lambda)$ of the 
form \eqref{eq:gindikin} on
the correspondence space $\CF$. 
}\vspace*{10pt}

\noindent
{\it Proof:} 
First, let us define a differential two-form $\Sigma^{AB}(\lambda)$ by
setting
$$\Sigma^{AB}(\lambda)\ :=\ \lambda_\ad\lambda_\bd\,E^{A\ad}\wedge E^{B\bd}.
$$
It then follows that $\dt_h\Sigma^{AB}$ is given by
$$
\dt_h\Sigma^{AB}\ =\ -2\lambda_\ad\lambda_\bd\, E^{[A\ad}\wedge\dt\, E^{B\}\bd},
$$
where $\dt$ is the exterior derivative on $\CM$. Assuming the
vanishing of the torsion and
upon substituting Eqs.\ \eqref{eq:deftor}
into this equation, we see that the connection one-form on $\CM$
will be of the form ${\Omega_{\ua}}^\ub={\Omega_{A\ad}}^{B\bd}
={\d_\ad}^\bd{\Omega_{A}}^B$ if and only if 
$$ 
\dt_h\Sigma^{AB}\ =\ -2\Sigma^{[AC}\wedge{\Omega_C}^{B\}}.
$$
Therefore, 
$$
\dt_h\Sigma\ =\ \dt_h(\Sigma^{AB}e_{BA})\ =\ 0,
$$
since $\dt e_{AB}-2{\Omega_{[A}}^Ce_{CB\}}=0.$

\hfill $\Box$ 

The differential two-form $\Sigma(\lambda)$ satisfying the properties
stated in the immediately preceding proposition is a
supersymmetric extension of the Gindikin two-form \cite{Gindikin}
(see also Ref.\ \cite{Plebanski:1975wn}). Note that the twistor distribution
$\Scr{D}_\CF=\langle\widetilde{E}_A\rangle$  annihilates $\Sigma(\lambda)$,
i.e. $\Sigma(\lambda)$ descends down to $\CP$ (also $\dt\Sigma(\lambda)$
is annihilated by $\Scr{D}_\CF$).

\paragraph{Supertwistor space for complex hyper-K\"ahler RC supermanifolds.}
The question which now arises is how the Gindikin two-form can be obtained
from certain data given on the supertwistor space $\CP$.
In the following, we generalize the results known from
the purely bosonic situation (see Penrose \cite{Penrose:1976js}
and also Alekseevsky and Graev \cite{Alekseevsky}). 

Let us assume that the supertwistor space $\CP$ is a holomorphic
fibre bundle $\pi\,:\CP\to\IP^1$ over the Riemann sphere $\IP^1$.
Later on, in \ref{par:HKSEB}~~ we shall see that this condition 
arises quite naturally. 
Furthermore, let us consider the line bundle $\CO_{\IP^1}(2)$ over
$\IP^1\hookrightarrow\CP$ together with its pull-back $\pi^*\CO_{\IP^1}(2)\to\CP$ 
to $\CP$.\footnote{Recall
that $\CO_{\IP^1}(m):=\CO_{\IP^1}(1)^{\otimes m}$.}
In addition, let $\Omega^1\CP/\IP^1$ be the sheaf of relative differential
one-forms on $\CP$ described by
\begin{equation}
 0\ \longrightarrow\ \pi^*\Omega^1\IP^1\ \longrightarrow\
 \Omega^1\CP\ \longrightarrow\ \Omega^1\CP/\IP^1\ \longrightarrow\ 0.
\end{equation}
According to Alekseevsky and Graev \cite{Alekseevsky}, we give the following
definition (already adopted to our situation): 

{\Def 
A section $\omega\in H^0(\CP,\Lambda^2(\Omega^1\CP/\IP^1)\otimes\pi^*\CO_{\IP^1}(2))$ 
of the sheaf $\Lambda^2(\Omega^1\CP/\IP^1)\otimes\pi^*\CO_{\IP^1}(2)$ is called
a holomorphic relative symplectic
structure of type $\CO_{\IP^1}(2)$ on $\CP$ if it is closed and non-degenerate on the
fibres. The integer ${\rm deg}(\CO_{\IP^1}(2))=c_1(\CO_{\IP^1}(2))=2$ is called the weight of $\omega$.\footnote{Notice that
for $\CN=0$ a relative differential two-form is automatically relatively closed as 
in this case the
fibres of $\pi$ are two-dimensional.}} \vspace*{10pt}

Then Gindikin's two-form $\Sigma(\lambda)$ on $\CF$
can be obtained by pulling back the relative symplectic structure
$\omega$ on $\CP$ to the correspondence space $\CF$ (and by dividing
it by a constant section of $\pi^*\CO_{\IP^1}(2)$). 

Altogether, we may now summarize all the findings from
 above by stating the following
theorem:

{\Thm\label{thm:NLG} 
There is a one-to-one correspondence between civilized 
 RC supermanifolds $\CM$ of dimension
$(4|2\CN)$ which are equipped with a hyper-K\"ahler structure and 
complex supermanifolds $\CP$ of dimension $(3|\CN)$ such that:
\begin{itemize}
\setlength{\itemsep}{-1mm}
\item[(i)] $\CP$ is a holomorphic fibre bundle $\pi\,:\,\CP\to\IP^1$
           over $\IP^1$,
\item[(ii)] $\CP$ is equipped with a $(4|2\CN)$-parameter family of
            sections of $\pi$, each with normal bundle given by \eqref{eq:normalbundle} and
\item[(iii)] there exists a holomorphic relative symplectic structure $\omega$ of
             weight $2$ on $\CP$.
\end{itemize}
}

\subsection{\kern-10pt Equivalent formulation ($\CN\neq4$)}\label{subsec:EF}

The purpose
of this section is to provide an alternative formulation of our
above considerations. In this way, we will also be able to
describe the case with nonzero cosmological constant. Here, we are generalizing
some of the results of Ward \cite{Ward:1980}, of LeBrun
\cite{LeBrun:1985,Lebrun:1986it}, 
of Bailey and Eastwood \cite{Bailey:1991} and of Merkulov 
\cite{Merkulov:1992b,Merkulov:1992qa}. 

Let $\CM$ be a civilized right-flat complex quaternionic RC supermanifold
with connection $\nabla$. Let further 
$\CP$ be its associated supertwistor space.
There exist several natural vector bundles
on $\CP$ 
which encode information about the supermanifold $\CM$
and about $\CP$ itself, respectively.
In the sequel, we shall be using the
notation
\begin{equation}\label{eq:PCW}
 \CE[m]\ :=\ \CE\otimes({\rm Ber}\,\widetilde\CS)^{-m}\ \cong\ \CE\otimes(\Lambda^2\widetilde\CS)^{-m}
\end{equation}
for any locally free sheaf $\CE$ on $\CM$.

\paragraph{Universal line bundle.}\label{par:ULB}
Let us begin by recalling Eq.\ \eqref{eq:lifteqre}.
In fact, this equation implies the existence of a natural holomorphic line 
bundle $\cL\to\CP$ over $\CP$. Following LeBrun's
terminology \cite{LeBrun:1985}, we shall refer to $\cL$ as the {\it universal
line bundle}. It is defined as follows.
Let $\CO_\CF(-1)$ be again the tautological bundle on
$\CF$. 
Furthermore, denote by $\Omega^1\CF/\CP$ the sheaf 
of relative differential one-forms on $\CF$ described by the sequence
\begin{equation}\label{eq:relone}
 0\ \longrightarrow\ \pi_2^*\Omega^1\CP\ \longrightarrow\
 \Omega^1\CF\ \longrightarrow\ \Omega^1\CF/\CP\ \longrightarrow\ 0.
\end{equation}
Then we may define the composition
\begin{equation}
 \nabla_{T\CF/\CP}\,:\,\CO_\CF(-1)\ \overset{\pi_1^*\nabla}{\longrightarrow}\
 \CO_\CF(-1)\otimes\pi_1^*\Omega^1\CM\ \overset{{\rm id}\otimes{\rm res}}{\longrightarrow}\
  \CO_\CF(-1)\otimes \Omega^1\CF/\CP,
\end{equation}
where  res denotes the restriction of 
differential one-forms on $\CF$ onto the fibres of the projection $\pi_2:\CF\to\CM$. 
The universal line bundle $\cL$ is then defined by the zeroth direct image
\begin{equation}
 \cL\ :=\ \pi_{2*}(\ker\nabla_{T\CF/\CP}).
\end{equation}
Hence, the fibre of $\cL$ over a point $z\in\CP$ is the space of solutions to 
$\lambda^\ad\nabla_{A\ad}\lambda_\bd=0$ on the $\beta$-surface 
$\pi_1(\pi_2^{-1}(z))$. 
Note that $\cL$ restricted to $\pi_2(\pi_1^{-1}(x))\hookrightarrow\CP$,  for $x\in\CM$,
can be identified with $\CO_{\IP^1}(-1)$.

\paragraph{Jacobi bundle.}\label{par:JB}
The second bundle over the supertwistor space
we are interested in is the so-called {\it Jacobi bundle}
(see also Refs.\ \cite{LeBrun:1985,Merkulov:1992qa}). Let us
denote it by $\Scr{J}$. It is defined to be the solution space of the
{\it supertwistor equation}
\begin{equation}\label{eq:sutw}
 \lambda^\ad(\nabla_{A\ad}\omega^B+{\d_A}^B\pi_\ad)\ =\ 0,
\end{equation}
on the $\beta$-surface $\pi_1(\pi_2^{-1}(z))$. Here,
$\lambda_\ad$ is non-zero and
obeys \eqref{eq:lifteqre} and $\pi_\ad$ is arbitrary. Note that
\eqref{eq:sutw} does not depend on the chosen scale on $\CM$. Note further that
the rank of $\Scr{J}$ is $3|\CN$. Then we have the following result:

{\Pro\label{pro:natiso}
 There is a natural isomorphism $T\CP\cong\Scr{J}\otimes \cL^{-1}$.}\vspace*{10pt}

\noindent
{\it Proof:} 
Let $z$ be a point in $\CP$ and $\Sigma:=\pi_1(\pi_2^{-1}(z))$ the associated
$\beta$-surface in $\CM$, and let $\lambda^\ad$ be a section of $\cL$. 

It is always possible to have a one-parameter foliation of $\Sigma$
since 
$$ \lambda_\bd\lambda^\ad\nabla_{A\ad}(\mu^B\lambda^\bd)
   \ =\ \lambda_\bd\lambda^\bd\lambda^\ad\nabla_{A\ad}\mu^B\ =\ 0, $$
where $\mu^A\lambda^\ad$ is any tangent vector field to $\Sigma$. Let now $J=J^{A\ad} E_{A\ad}$
be the associated Jacobi field on $\Sigma$. Then any tangent vector $(\omega^A,\pi_\ad)$ at 
$z\in\CP$ can be represented by Jacobi fields on $\Sigma$,
$$ \omega^A\ =\ J^{A\ad}\lambda_\ad\qquad{\rm and}\qquad \pi_\ad\ =\ 
J^{B\bd}\nabla_{B\bd}\lambda_\ad,$$  
subject to 
the constraint 
$$ \mathcal{L}_JX\ {\rm mod}\ T\Sigma\ =\ 
 [J,X\}\ {\rm mod}\ T\Sigma\ =\ 0\qquad{\rm for\ any}\qquad X\in T\Sigma. $$
Note that the above equations are unaffected by changes of the form $J^{A\ad}\mapsto J^{A\ad}+
J^A\lambda^\ad$, where $J^A\lambda^\ad$ is also a Jacobi field which is in addition 
tangent to $\Sigma$. Therefore, tangent vectors at $z\in\CP$
are actually represented by {\it equivalence classes}
of Jacobi fields, where two Jacobi fields
are said to be equivalent if their difference lies in $T\Sigma$. 

Explicitly, the constraint $[J,X\}\in T\Sigma$ reads as
\begin{equation}\label{eq:JE}
\lambda_\bd(\lambda^\ad\nabla_{A\ad}J^{B\bd}-{\d_A}^BJ^{C\gd}\nabla_{C\gd}\lambda^\bd)\ =\ 0.
\end{equation}
Using this expression, one may straightforwardly check that 
$\omega^A$ obeys the supertwistor equation \eqref{eq:sutw}. Hence, the mapping
$J\otimes\lambda\mapsto\omega$ defines a morphism $T\CP\otimes\cL\to\Scr{J}$.
Since the solution space of \eqref{eq:sutw} is of
the right dimensionality, we have thus constructed an
isomorphism.

\hfill $\Box$

\paragraph{Einstein bundle.}
The last vector bundle we are about to define is the {\it Einstein bundle}. Originally,
it was introduced by LeBrun \cite{LeBrun:1985} in the context of the ambitwistor space
(the space of complex null-geodesics of some given complex four-dimensional manifold)
and its relation to the (full) Einstein equations. He showed that non-vanishing sections of this 
bundle are in one-to-one correspondence with Einstein metrics in the given conformal class.
Unfortunately, the Einstein bundle on ambitwistor space and its generalization to superambitwistor
space in the context of $\CN=1$ supergravity (cf. Merkulov \cite{Merkulov:1992qa}) 
seem only to be definable in terms of their inverse images on the associated
correspondence space, that is, so far it lacks a description in terms of the 
intrinsic structure of the (super)ambitwistor space. As we shall see in a
moment, this will not be the case if the (super)manifold under consideration is
(super)conformally right-flat. It is this additional condition that allows for 
giving an explicit description of this bundle in terms of natural holomorphic sheaves
on the (super)twistor space. As we shall see, this bundle will also 
yield a reinterpretation of the results given in Thm.\ 
\ref{thm:NLG} Our subsequent discussion is a generalization of the ideas of
\cite{LeBrun:1985,Bailey:1991,Merkulov:1992b,Merkulov:1992qa}.

Next we introduce a second-order differential operator, $\Delta$,
on the correspondence space $\CF$ which is given in the structure frame by
\begin{equation}
 \Delta_{AB}\ :=\ \lambda^\ad\lambda^\bd(\nabla_{\{A\ad}\nabla_{B]\bd}+R_{AB\ad\bd}),
\end{equation}
where $\lambda_\ad$ obeys \eqref{eq:lifteqre} and $R_{AB\ad\bd}$ is the $R^+_{\widetilde\CS}$-part 
of the curvature
as discussed in Prop.\ \ref{pro:curv} 
It then follows that $\Delta$ is independent of
the choice of scale if it acts on sections of $\pi_1^{-1}\CO_\CM[-1]$.
This can be seen as follows: let $\varphi$ be a section of $\CO_\CM[k]$. If
one performs a change of scale according to $\tilde\varepsilon\mapsto\g\tilde\varepsilon$,
the connection changes as follows:
\begin{equation}
 \widehat\nabla_{\widehat A\widehat\ad}\varphi\ =\ 
 \g^{-\kappa}(\nabla_{A\ad}\varphi-k\g_{A\ad}\varphi),
\end{equation}
where $\kappa$ was defined in \eqref{eq:defkap}
and $\g_{A\ad}:=E_{A\ad}\log\g$, as before. Therefore, if one chooses $k=-1$ one
arrives after a few lines of algebra at
\begin{equation}
 \widehat\nabla_{\{\widehat A(\widehat\ad}\widehat\nabla_{\widehat B]
   \widehat\bd)}\varphi\ =\ \g^{-2\kappa}(
 \nabla_{\{A(\ad}\nabla_{B]\bd)}\varphi+\nabla_{\{A(\ad}\g_{B]\bd)}\varphi-
    \g_{\{A(\ad}\g_{B]\bd)}\varphi).
\end{equation}
In a similar manner, one may verify that
\begin{equation}
 \widehat{R}_{\widehat A\widehat B\widehat\ad\widehat\bd}\ =\ 
   \g^{-2\kappa}(R_{AB\ad\bd}-\nabla_{\{A(\ad}\g_{B]\bd)}+
    \g_{\{A(\ad}\g_{B]\bd)}).
\end{equation}
Combining these two expressions, one arrives at the desired result. 

As $\Delta$ acts on the fibres of $\pi_2\,:\,\CF\to\CP$, we can define the
Einstein bundle $\CE$ on $\CP$ by the following resolution:
\begin{equation}\label{eq:EBR}
 0\ \longrightarrow\ \pi_2^{-1}\CE\ \longrightarrow\ \pi_1^{-1}\CO_\CM[-1]\
 \overset{\Delta}{\longrightarrow}\ \pi_1^*(\odot^2\CH^\vee[-3])\otimes\CO_\CF(2)
  \ \longrightarrow\ 0,
\end{equation}
where $\CO_\CF(2)$ is the second tensor power of the dual of the tautological
bundle $\CO_\CF(-1)$ on the correspondence space $\CF$.

We are now in the position to relate the four bundles $T\CP$, $\Scr{L}$, $\Scr{J}$ and
$\CE$ among themselves by virtue of the following proposition:

{\Pro\label{pro:EBdec}
 There is a natural isomorphism of sheaves $\CE\cong \Omega^1\CP\otimes\cL^{-2}\cong
\Scr{J}^\vee\otimes\Scr{L}^{-1}$.}\vspace*{10pt}

\noindent
{\it Proof:}
The second isomorphism is the one proven in Prop.\ \ref{pro:natiso} So it 
remains to verify the first one.
Recall again that $T\CP\cong\Scr{J}\otimes\cL^{-1}$, that is, the fibre
of $T\CP$ over some point $z\in\CP$ is the space of solutions of the
supertwistor equation on $\pi_2^{-1}(z)$ for $\omega^A$ being of homogeneous degree one in
$\lambda_\ad$. The fibre of the Einstein bundle $\CE$ over $z\in\CP$ coincides
with the kernel of $\Delta$ on the same subsupermanifold $\pi_2^{-1}(z)\hookrightarrow\CF$. 

Consider now the scalar
$$
 Q\ :=\ (2-\CN)\omega^A\lambda^\ad\nabla_{A\ad}\varphi-\varphi(-)^{p_A}\lambda^\ad\nabla_{A\ad}\omega^A,
$$
where $\varphi$ is a section of $\pi^{-1}_1\CO_\CM[-1]$ and $\omega^A$
a solution to the supertwistor equation \eqref{eq:sutw}. Clearly, $Q$ is of homogeneous degree two
in $\lambda_\ad$ and as one may check, it is independent of the choice of scale. In
showing the latter statement, one needs the relation
$$
 \lambda^\ad\nabla_{A\ad}\omega^B\ =\ \tfrac{1}{2-\CN}{\d_A}^B(-)^{p_C}\lambda^\gd\nabla_{C\gd}\omega^C,
$$
which follows from the supertwistor equation \eqref{eq:sutw}.
In addition, upon using the very same equation \eqref{eq:sutw}
together with $\Delta_{AB}\varphi=0$, one finds that
$$
 \lambda^\ad\nabla_{A\ad} Q\ =\ 0.
$$ 
Hence, the quantity $Q$ corresponds to a point in the fibre of $\cL^{-2}$ over the
point $z\in\CP$. Altogether, $Q$ provides a non-degenerate $\cL^{-2}$-valued pairing of the fibres
of tangent bundle $T\CP$ and of the Einstein bundle $\CE$, thus establishing the
claimed isomorphism.

\hfill $\Box$ 

\noindent
This shows, as indicated earlier,
 that the Einstein bundle is fully determined in terms of the intrinsic
structure of the supertwistor space.

\paragraph{Hyper-K\"ahler structures.}\label{par:HKSEB}
The next step is to verify the following statement:

{\Pro There is a natural one-to-one correspondence between scales on a
civilized right-flat complex quaternionic RC supermanifold $\CM$ in
which the $R_{\widetilde\CS}^+$-part of the curvature vanishes and nonvanishing sections of the Einstein
bundle $\CE$ over the associated supertwistor space $\CP$.}\vspace*{10pt}

\noindent
Putting it differently, nonvanishing sections of the Einstein bundle are
in one-to-one correspondence with (equivalence classes of) solutions to
 the self-dual supergravity
equations with nonzero cosmological constant.

\noindent
{\it Proof:}
By our convexity assumption
(recall that $\CM$ is assumed to be civilized; putting it differently, there is 
a Stein covering
of $\CM$), we have $H^q(\pi_2^{-1}(z),\IC)\cong0$
for $z\in\CP$ and $q\geq1$. Let $\CU\subset\CM$ be an open subset 
and set $\CU':=\pi_1^{-1}(\CU)\subset \CF$ and $\CU'':=\pi_2(\CU')\subset\CP$.
Therefore, we have an isomorphism\footnote{Note 
that this in fact holds true for any
locally free sheaf on the supertwistor space.}
$$
 H^r(\CU'',\CE)\ \cong\ H^r(\CU',\pi_2^{-1}\CE).
$$ 
Hence, in order to compute $H^r(\CU'',\CE)$ we need to compute $H^r(\CU',\pi_2^{-1}\CE)$.
However, the latter cohomology groups can be computed from the exact 
resolution \eqref{eq:EBR} upon applying the direct image functor
$$
    0\ \longrightarrow\ \pi_{1*}(\pi_2^{-1}\CE)\ \longrightarrow\
  \pi_{1*}\CR^0\ \longrightarrow\ 
          \pi_{1*}\CR^1\ \longrightarrow\ 
          \pi_{1*}^1(\pi_2^{-1}\CE)
$$ 
where we have abbreviated
$$
 \CR^0\ :=\ \pi_1^{-1}\CO_\CM[-1]\qquad{\rm and}\qquad \CR^1\ :=\ 
 \pi_1^*(\odot^2\CH^\vee[-3])\otimes\CO_\CF(2).
$$
In addition, there is a spectral sequence converging to 
$$ H^{p+q}(\CU', \pi_2^{-1}\CE),$$
with
$$ E_1^{p,q}\ \cong\ H^0(\CU,\pi_{1*}^q\CR^p). $$
Notice the sheaves in resolution have vanishing higher
direct images while the zeroth images are given by 
$$
\begin{aligned}
 \pi_{1*}\CR^0\ &\cong\ \CO_\CM[-1],\\
 \pi_{1*}\CR^1\ &\cong\ (\odot^2\CH^\vee\otimes\odot^2\widetilde\CS^\vee)[-1].
\end{aligned}
$$
Therefore, the cohomology group
$H^0(\CU'',\CE)\cong H^0(\CU',\pi_2^{-1}\CE)$ is isomorphic
to the kernel of a second order differential operator which by virtue of our
above discussion turns out to be the solution space of 
$$
(\nabla_{\{A(\ad}\nabla_{B]\bd)}+R_{AB\ad\bd})\varphi\ =\ 0,
$$
where $\varphi$ is a nonvanishing section of $\CO_\CM[-1]$. As discussed
above, this equation is independent of the choice of scale on $\CM$. 
Since $\varphi$ is a nonvanishing section of $\CO_\CM[-1]$, we may always
work in the scale where $\varphi=1$. Thus, the above equation implies
that $R_{AB\ad\bd}$ must vanish and the proof is completed.

\hfill $\Box$

Let now $\tau$ be the section of $\CE$ corresponding to the
scale where $R^+_{\widetilde\CS}$ vanishes. Obviously, it defines a
$(2|\CN)$-dimensional distribution on the supertwistor space
$\CP$ given by the kernel of $\tau$. One also says that $\tau$
is a non-degenerate holomorphic contact form determing a holomorphic 
contact structure (a distribution of Gra{\ss}mann 
even codimension one) on $\CP$. Thus, 
non-degeneracy of $\tau$ insures a nonvanishing cosmological
constant. 
In addition, let $\nabla$ be the connection on $\CM$ defined
by this chosen scale. 
In the remainder, we shall show that degenerate contact
structures on $\CP$ are in one-to-one correspondence with
(equivalence classes of) solutions to the self-dual supergravity equations with
zero cosmological constant. 

{\Pro\label{prop:equivalence} 
The $(4|2\CN)$-dimensional distribution on $\CF$ defined by $\pi_2^*\tau$ 
coincides with the $(4|2\CN)$-dimensional distribution defined by
$\pi_1^*\nabla$.
}\vspace*{10pt}

\noindent
{\it Proof:} Recall that $\CE\cong \Omega^1\CP\otimes \cL^{-2}$. Then we note that
the pairing $\Omega^1\CP\otimes\cL^{-2}\times T\CP\to \cL^{-2}$ is given by
$$
(2-\CN)\omega^A\lambda^\ad\nabla_{A\ad}\varphi-\varphi(-)^{p_A}\lambda^\ad\nabla_{A\ad}\omega^A,
$$
as follows by the discussion given in the proof of Prop.\ \ref{pro:EBdec}
Here, $\varphi$ represents $\tau$ on $\CF$ and $\omega^A$ corresponds to a
tangent vector on $\CP$. In the scale defined by $\tau$, we have $\varphi=1$
(cf. the proof of the immediately preceding proposition). By virtue of the
the supertwistor equation \eqref{eq:sutw}, we conclude that the distribution
on the correspondence space $\CF$ defined by the vanishing of this pairing is given by
$$
 \lambda^\ad\nabla_{A\ad}\omega^B\ =\ 0.
$$
Eq.\ \eqref{eq:JE} in turn implies that a solution to this equation must correspond to 
a Jacobi field $J^{A\ad}$ which satisfies
$$ J^{A\ad}\lambda^\bd \nabla_{A\ad}\lambda_\bd\ =\ 0.
$$
Therefore, we have a correspondence between subspaces of the fibre of $T\CP$ over a point $z\in\CP$
which are annihilated by the differential one-form $\tau$ and Jacobi
fields on the $\beta$-surface $\Sigma=\pi_1(\pi_2^{-1}(z))\hookrightarrow\CM$
which are annihilated by the differential one-form
$
 E^{A\ad}\lambda^\bd\nabla_{A\ad}\lambda_\bd.
$
In fact, this form is the push-forward to $\CM$ of the differential one-form on $\CF$
defining the distribution given by $\pi_1^*\nabla$.

\hfill $\Box$

Then we have the following result:

{\Pro Let $\tau$ be the section of the Einstein bundle $\CE\to\CP$ corresponding to 
the scale in which $R^+_{\widetilde\CS}$ vanishes. In this scale, the cosmological
constant will vanish if and only if the distribution on $\CP$ defined by $\tau$ is
integrable. Hence, the Ricci tensor is zero.}\vspace*{10pt}

\noindent
This means that nonvanishing integrable sections of the Einstein bundle
(that is, degenerate contact structures)
are in one-to-one correspondence with (equivalence classes of) solutions to
 the self-dual supergravity equations
with zero cosmological constant.

\noindent
{\it Proof:}
Obviously, showing integrability of the distribution on $\CP$ defined by $\tau$ is
equivalent to showing the integrability of the distribution on $\CF$ defined by the
pull-pack $\pi_2^*\tau$. Prop.\ \ref{prop:equivalence} implies that the distribution
defined by $\pi_2^*\tau$ will be integrable if and only if $\widetilde\CS^\vee$ is
projectively flat in the scale defined by $\tau$ (see also comment after proof of
Prop.\ \ref{pro:PRF} leading to Eq.\ \eqref{eq:lifteqre}). 
We conclude from Prop.\ \ref{pro:curv} 
that in addition to $R_{AB\ad\bd}$ also $\Lambda_{AB}$ must vanish.
Hence, the Ricci tensor is zero.

\hfill $\Box$

Recall that $\tau$ defines a $(2|\CN)$-dimensional distribution on $\CP$. If
this distribution is integrable, it gives a foliation of $\CP$ by $(2|\CN)$-dimensional
subsupermanifolds. In fact, it yields a holomorphic fibration
\begin{equation}
 \CP\ \to\ \IP^1
\end{equation}
of the supertwistor space over the Riemann sphere (see also Penrose
\cite{Penrose:1976js} for the purely bosonic situation).
Remember that this fibration was 
one of the assumptions made in Thm.\ \ref{thm:NLG} Therefore, we may conclude that
if the distribution $\tau$ is integrable the supertwistor space
$\CP$ is equipped with a relative symplectic structure as stated in point (iii)
of Thm.\ \ref{thm:NLG}

\paragraph{Summary.}
Let us summarize all the correspondences derived above in the following table:

\begin{center}
\let\PBS=\PreserveBackslash
\setlength{\extrarowheight}{4pt}
 \begin{tabular}{| >{\PBS\raggedright\hspace{0pt}}p{2in} 
                 | >{}p{3.5in}| }  
 \hline
      supertwistor spaces $\CP$              
        & civilized right-flat complex quaternionic RC supermanifolds, i.e.
           ${C_{AB\ad\bd\gd}}^\dd=0$ \\
\hline
      supertwistor spaces $\CP$ with non-degenerate holomorphic contact structures
        & civilized right-flat complex quaternionic RC supermanifolds which are
          self-dual Einstein, i.e.
           ${C_{AB\ad\bd\gd}}^\dd=0$ and ${R_{AB\ad}}^\bd=0$\\
\hline
     supertwistor spaces $\CP$ with degenerate holomorphic contact structures
        & civilized right-flat complex quaternionic RC supermanifolds which are
          self-dual, i.e.
           ${C_{AB\ad\bd\gd}}^\dd=0$, ${R_{AB\ad}}^\bd=0$ and $\Lambda_{AB}=0$\\
\hline
 \end{tabular}
\end{center}
We remind the reader that the curvature components can be found in Prop.\ \ref{pro:curv}

\subsection{\kern-10pt Bundle of local supertwistors ($\CN\neq4$)}

This subsection is devoted to the bundle of local supertwistors and its
implications on the supermanifolds under consideration. Here, we give
a generalization of methods developed by Penrose \cite{Penrose}, by
LeBrun \cite{LeBrun:1985} and by Bailey and Eastwood \cite{Bailey:1991}.
So, let $\CM$ be a civilized right-flat complex quaternionic RC supermanifold
with connection $\nabla$, in the sequel.

\paragraph{Bundle of local supertwistors.}\label{par:BoLST}
Let us start by recalling the jet
sequence (for a proof, see e.g. Manin \cite{Manin})
\begin{equation}\label{eq:jetseq}
 0\ \longrightarrow\ 
 \Omega^1\CM\otimes \CE\ \longrightarrow\ {\rm Jet}^1\CE\ \longrightarrow\ \CE\ \longrightarrow\ 0,
\end{equation}
where $\CE$ is some locally free sheaf on $\CM$ and
${\rm Jet}^1\CE$ is the sheaf of first-order jets of $\CE$. Recall further
the factorization of the tangent bundle of $\CM$ as $T\CM\cong\CH\otimes\widetilde\CS$. Choose 
now $\CE$ to be $\CH$. Hence, the above sequence becomes
\begin{equation}\label{eq:LSTB1}
 0\ \longrightarrow\ 
 (\CH\otimes \CH^\vee)\otimes\widetilde\CS^\vee\ 
 \longrightarrow\ {\rm Jet}^1\CH\ \longrightarrow\ \CH\ \longrightarrow\ 0.
\end{equation}
Since, $(\CH\otimes \CH^\vee)_0\otimes\widetilde\CS^\vee$, where 
$(\CH\otimes \CH^\vee)_0$ means the trace-free part of $\CH\otimes \CH^\vee$,
is a subbundle of 
$\Omega^1\CM\otimes\CH$, i.e.
\begin{equation}
 0\ \longrightarrow\ 
 (\CH\otimes \CH^\vee)_0\otimes\widetilde\CS^\vee\ 
 \longrightarrow\ \Omega^1\CM\otimes\CH,
\end{equation}
we may define a rank-$4|\CN$ bundle, denoted by $\Scr{T}$, over $\CM$ by
the following sequence:
\begin{equation}\label{eq:LSTB2}
 0\ \longrightarrow\ 
 (\CH\otimes \CH^\vee)_0\otimes\widetilde\CS^\vee\ 
 \longrightarrow\ {\rm Jet}^1\CH\ \longrightarrow\ \Scr{T}\ \longrightarrow\ 0.
\end{equation}
We shall call $\Scr{T}$ the {\it bundle of local supertwistors}.
The reason for naming it like this will become clear in due course of
our subsequent discussion.

As a first result, we obtain from \eqref{eq:LSTB1} and \eqref{eq:LSTB2} a
natural isomorphism:
\begin{equation}
 {\rm Ber}\,\Scr{T}\ \cong\ {\rm Ber}\,\CH\otimes({\rm Ber}\,\widetilde\CS)^{-1}.
\end{equation}
Hence, by virtue of \eqref{eq:beriso} we may conclude that
\begin{equation}\label{eq:berLSTB}
 {\rm Ber}\,\Scr{T}\ \cong\ \CO_\CM.
\end{equation}
Furthermore, in a structure frame, $\Scr{T}$ may be described by 
natural fibre coordinates of the form $(\omega^A,\pi_\ad)$.
Under a change of scale $\tilde\varepsilon\mapsto\g\tilde\varepsilon$, 
these coordinates behave as
\begin{equation}\label{eq:scalingLST}
 \omega^A\ \mapsto\ \widehat\omega^{\widehat A}\ =\ \g^{\frac{1}{2}\kappa}\omega^A
 \qquad{\rm and}\qquad
 \pi_\ad\ \mapsto\ \widehat\pi_{\widehat\ad}\ =\ \g^{-\frac{1}{2}\kappa}(\pi_\ad-\omega^A\g_{A\ad}),
\end{equation}
which is an immediate consequence of the transformation laws \eqref{eq:TLCRL}.
Remember that the constant $\kappa$ appearing above was introduced in 
\eqref{eq:defkap} and $\g_{A\ad}$ was defined to be
$\g_{A\ad}=E_{A\ad}\log\g$. Altogether, these considerations
imply that there is a canonical exact sequence
\begin{equation}
 0\ \longrightarrow\ 
   \widetilde\CS^\vee\ \longrightarrow\ \Scr{T}\ \longrightarrow\ \CH\ \longrightarrow\ 0.
\end{equation}
 
\paragraph{Local supertwistor connection.}\label{par:LSC}
In the class of affine connections on the bundle
$\Scr{T}$, there exists a distinguished one referred to as
the {\it local supertwistor connection}, in the following. 
This is an immediate consequence of the
scaling behavior \eqref{eq:scalingLST}, as we shall see
now. Let us mention in passing that this particular connection
will be unique and independent of the choice of scale on $\CM$.

Let us recall the supertwistor
equation \eqref{eq:sutw}, which we repeat for the 
reader's convenience at this stage
\begin{equation}\label{eq:LST1}
 \lambda^\ad(\nabla_{A\ad}\omega^B+{\d_A}^B\pi_\ad)\ =\ 0.
\end{equation}
Recall further from the proof of Prop.\ \ref{pro:natiso} that
tangent vectors at $z\in\CP$  
can be represented by $\omega^A=J^{A\ad}\lambda_\ad$ and 
$\pi_\ad=J^{B\bd}\nabla_{B\bd}\lambda_\ad$, where
$J=J^{A\ad}E_{A\ad}$ is a Jacobi field on the $\beta$-surface
$\pi_1(\pi_2^{-1}(z))\hookrightarrow\CM$. 
In the very same 
proof, we have argued that this $\omega^A$ satisfies the
supertwistor equation. Similarly, one may show that
\begin{equation}\label{eq:LST2}
 \lambda^\ad(\nabla_{A\ad}\pi_\bd+(-)^{p_B}(R_{AB\ad\bd}-\Lambda_{AB}\epsilon_{\ad\bd})\omega^B)\ =\ 0.
\end{equation}
Here, we have made use of the curvature decompositions
 \eqref{eq:procurv}. Furthermore, the
scaling behavior \eqref{eq:scalingLST} is exactly of the 
same form as the one of $\omega^A=J^{A\ad}\lambda_\ad$ and 
$\pi_\ad=J^{B\bd}\nabla_{B\bd}\lambda_\ad$, respectively. That it is why
we have denoted the fibre coordinates
of the bundle $\Scr{T}$ by the same letters.

Altogether, Eqs.\ \eqref{eq:LST1} and \eqref{eq:LST2} can be reinterpreted
as an $SL(4|\CN)$-connection $\mathcal{D}$ on $\Scr{T}$ -- the local supertwistor connection:
\begin{equation}\label{eq:SC1}
 \lambda^\ad\mathcal{D}_{A\ad}\binom{\omega^B}{\pi_\bd}\ =\  
 \binom{\lambda^\ad(\nabla_{A\ad}\omega^B+{\d_A}^B\pi_\ad)}{\lambda^\ad(
  \nabla_{A\ad}\pi_\bd+(-)^{p_B}(R_{AB\ad\bd}-\Lambda_{AB}\epsilon_{\ad\bd})\omega^B)}.
\end{equation}
Thus, flat sections with respect to this connection correspond
to solutions of the supertwistor equation. After all, this justifies the
name local supertwistor bundle.

Let us make the following abbreviations: $\mathcal{D}_A:=\lambda^\ad\mathcal{D}_{A\ad}$,
$\nabla_A:=\lambda^\ad\nabla_{A\ad}$ and 
$^t Z:=(\omega^A,\pi_\ad)$. Then we may rewrite Eqs.\ \eqref{eq:SC1}
concisely as
\begin{equation}
 \mathcal{D}_A Z\ =\ \nabla_A Z +\mathcal{A}_A Z,
\end{equation}
where 
\begin{equation}
 \mathcal{A}_A\ :=\ \begin{pmatrix}
                       0 & {\d_A}^B\lambda^\bd\\
                       (-)^{p_B}\lambda^\ad(R_{AB\ad\bd}-\Lambda_{AB}\epsilon_{\ad\bd}) & 0 
                    \end{pmatrix},
\end{equation}
is the $\mathfrak{sl}(4|\CN)$-valued gauge potential.
The local supertwistor connection $\mathcal{D}$ is torsion-free, since $\nabla$ is torsion-free.
Furthermore, the $\mathcal{F}^-$-part of the curvature two-form 
$\mathcal{F}=\mathcal{D}^2=\mathcal{F}^-+\mathcal{F}^+$ of $\mathcal{D}$ 
(here, we are using the notation of Prop.\ \ref{pro:curv}) is given in
a structure frame by
\begin{equation}
\begin{aligned}
 \mathcal{F}_{AB}\ &=\ \lambda^\ad\lambda^\bd\mathcal{F}_{A\ad B\bd}\\
                   &=\ [\mathcal{D}_A,\mathcal{D}_B\}\\
                   &=\ R_{AB}+\nabla_A\mathcal{A}_B-(-)^{p_Ap_B}\nabla_B\mathcal{A}_A+
                        [\mathcal{A}_A,\mathcal{A}_B\},
\end{aligned}
\end{equation}
where $R_{AB}=[\nabla_A,\nabla_B\}=\lambda^\ad\lambda^\bd[\nabla_{A\ad},\nabla_{B\bd}\}$.
Next one verifies that $\nabla_{[A}\mathcal{A}_{B\}}$ actually vanishes,
which is due to Eqs.\ \eqref{eq:lifteqre} and due to Bianchi identities
of the curvature of $\nabla$. Upon explicitly computing the commutator $[\mathcal{A}_A,\mathcal{A}_B\}$
and upon comparing it with $R_{AB}$ thereby using Eqs.\ \eqref{eq:procurv}, one
realizes that $R_{AB}=-[\mathcal{A}_A,\mathcal{A}_B\}$. In showing this,
one also needs to use the property that $\CM$ is right-flat. Therefore, we have
verified the following fact:

{\Pro The $\mathcal{F}^-$-part of the curvature $\mathcal{F}$ of the local supertwistor connection
$\mathcal{D}$ on the local supertwistor bundle $\Scr{T}$ over 
a civilized right-flat complex quaternionic RC supermanifold $\CM$ is 
zero. Hence, the curvature $\mathcal{F}$ is self-dual, that is, $\mathcal{F}=\mathcal{F}^+$.
}\vspace*{10pt}

\noindent
Putting it differently, the connection $\mathcal{D}$ is flat on any $\beta$-surface
$\pi_1(\pi_2^{-1}(z))\hookrightarrow\CM$ for all $z\in\CP$.
This is going to be important in the paragraph subsequent to the following
one, where we will show that the bundle of first-order jets of the dual 
universal line bundle $\Scr{L}$ over the supertwistor
space, i.e. Jet$^1\Scr{L}^{-1}\to\CP$,  corresponds to the dual of the
bundle of 
local supertwistors $\Scr{T}^\vee\to\CM$ by means of the
Penrose-Ward transform.

\paragraph{Penrose-Ward transform.}
In this paragraph, we briefly discuss the general form of the
Penrose-Ward transform which relates
certain holomorphic vector bundles over the supertwistor space
$\CP$ to holomorphic vector bundles over $\CM$ and vice versa. 
However, we
merely quote the result. A detailed proof goes along
the lines presented by Manin \cite{Manin} and can be done
in the supersymmetric setting without difficulties.
  
Suppose we are given a locally free sheaf $\CE_\CP$ on
$\CP$. Suppose further that $\CE_\CP$ is free when restricted to
any submanifold $\pi_2(\pi_1^{-1}(x))\hookrightarrow\CP$ for all $x\in\CM$. 
In addition, let $\Omega^1\CF/\CP$ the sheaf of relative 
differential one-forms on $\CF$ as given by the sequence
\eqref{eq:relone}. Furthermore, let
\begin{equation}
 \mathcal{D}_{T\CF/\CP}\,:\,\pi_2^*\CE_\CP\ \longrightarrow\
 \pi_2^*\CE_\CP\otimes\pi_1^*\Omega^1\CM\ \overset{{\rm id}\otimes{\rm res}}{\longrightarrow}\
  \pi_2^*\CE_\CP\otimes \Omega^1\CF/\CP
\end{equation}
be the relative connection on the pull-back $\pi_2^*\CE_\CP$
of $\CE_\CP$ to the correspondence space $\CF$.

In order for the below theorem to work, one needs
\begin{equation}
 \Omega^1\CM\ \cong\ \pi_{1*}\Omega^1\CF/\CP,
\end{equation}
since only then $\mathcal{D}_{T\CF/\CP}$ gives rise to
a connection $\mathcal{D}:=\pi_{1*}(\mathcal{D}_{T\CF/\CP})$ 
on $\CE_\CM=\pi_{1*}(\pi_2^*\CE_\CP)$.
One may check that this isomorphism indeed
follows from the sequence \eqref{eq:aseq2} after
dualizing and upon applying the direct image functor. In
showing this, one uses the fact that the direct images 
$\pi_{1*}(\pi^*_2\cN^\vee_{\IP^1|\CP})$ and
$\pi^1_{1*}(\pi^*_2\cN^\vee_{\IP^1|\CP})$ vanish due to
Serre duality. Since 
the fibres of $\pi_1\,:\,\CF\to\CM$ are compact and connected and
the ones of $\pi_2\,:\,\CF\to\CP$ are connected and simply connected
(recall that $\CM$ is assumed to be civilized), we have the following
theorem:

{\Thm\label{thm:PWT} 
Let $\CM$ be a civilized right-flat complex quaternionic RC
supermanifold and $\CP$ its associated supertwistor space. Then there
is a natural one-to-one correspondence between: 
\begin{itemize}
\setlength{\itemsep}{-1mm}
\item[(i)] the category of locally free sheaves $\CE_\CP$ on $\CP$
           which are free on any submanifold\linebreak[4] 
           $\pi_2(\pi_1^{-1}(x))\hookrightarrow\CP$ for all $x\in\CM$ and
\item[(ii)] the category of pairs $(\CE_\CM,\mathcal{D})$, where $\CE_\CM$ is a locally
           free sheaf on $\CM$ given by $\CE_\CM=\pi_{1*}(\pi_2^*\CE_\CP)$ and
           $\mathcal{D}$ is the push-forward of the relative connection on $\CF$, i.e.
           $\mathcal{D}:=\pi_{1*}(\mathcal{D}_{T\CF/\CP})$ which is flat on any $\beta$-surface
           $\pi_1(\pi_2^{-1}(z))\hookrightarrow\CM$ for all $z\in\CP$.
\end{itemize}
}\vspace*{10pt}

\noindent
Notice that flatness on any $\beta$-surface is equivalent to saying that
the curvature of $\mathcal{D}$ is self-dual. The above correspondence is
called {\it Penrose-Ward transform}.

\paragraph{Penrose-Ward transform of $\Scr{T}$.}
Consider the bundle of local supertwistors $\Scr{T}$ as defined
in \ref{par:BoLST}.~ As we have shown in \ref{par:LSC}~~, the local
supertwistor connection is self-dual, i.e.
flat on any $\beta$-surface in $\CM$. So one naturally asks for the
Penrose-Ward transform of $\Scr{T}$. The answer
gives the following proposition:

{\Pro The Penrose-Ward transform takes the bundle of local supertwistors
$\Scr{T}$ over $\CM$ to the dual of the
sheaf of first-order jets ${\rm Jet}^1\Scr{L}^{-1}$ of the dual
universal line bundle $\Scr{L}$ over $\CP$.}\vspace*{10pt}

\noindent
{\it Proof:} As a first check, notice that the restriction of $\Scr{L}$ to any 
$\pi_2(\pi_1^{-1}(x))\hookrightarrow\CP$ is $\CO_{\IP^1}(-1)$. Hence,
the dual restricts to $\CO_{\IP^1}(1)$. Hence, one may check that the dual of the
sheaf of first-order jets $({\rm Jet}^1\Scr{L}^{-1})^\vee$ of $\Scr{L}^{-1}$
is free when restricted to $\pi_2(\pi_1^{-1}(x))\hookrightarrow\CP$
as a consequence of the Euler sequence.\footnote{Recall that the
Euler sequence is given by:
$0\longrightarrow\CO_{\IP^n}\longrightarrow\CO_{\IP^n}(1)\otimes\IC^{n+1}\longrightarrow
T\IP^n\longrightarrow0$.} So, $({\rm Jet}^1\Scr{L}^{-1})^\vee$ satisfies point (i) of Thm.
\ref{thm:PWT} 

In the following, we are using a supersymmetric generalization
of an argument by LeBrun \cite{Lebrun:1986it}.
Let $m_t\,:\,\Scr{L}\setminus\{0\}\to\Scr{L}\setminus\{0\}$ (zero section
deleted), where 
$t\in\IC\setminus\{0\}$, denote the scalar multiplication map. Furthermore,
let $m_{t*}\,:\,T\Scr{L}\to T\Scr{L}$ be its Jacobian. 
According to LeBrun \cite{Lebrun:1986it}, one has an isomorphism
$$
 {\rm Jet}^1\Scr{L}^{-1}\ \cong\ (\Scr{L}\otimes(T\Scr{L}/\{m_{t*}\}))^\vee.
$$ 
Thus, we are about to verify that
$$
 \Scr{T}\ \cong\ \pi_{1*}(\pi_2^*(\Scr{L}\otimes(T\Scr{L}/\{m_{t*}\}))).
$$
To do this, we first recall that a point $\ell$ of $\Scr{L}$ is a pair
$(\pi_1(\pi_2^{-1}(z)),\lambda_\ad)$, where $z\in\CP$ and $\lambda_\ad$ is an
auto-parallel tangent spinor, i.e. it satisfies
Eq.\ \eqref{eq:lifteqre}. Therefore, a tangent vector at $\ell\in\Scr{L}$
can be represented by Jacobi fields as introduced and discussed in
the proof of Prop.\ \ref{pro:natiso} In particular, we may write
$$(\omega^A,\pi_\ad)\ =\ (J^{A\bd}\lambda_\bd,J^{B\bd}\nabla_{B\bd}\lambda_\ad),$$
for the tangent vector at $\ell\in\Scr{L}$. From our dicussion given 
in \ref{par:LSC}~\,\,, we know that such $(\omega^A,\pi_\ad)$
satisfy
$$
\begin{aligned}
 \lambda^\ad(\nabla_{A\ad}\omega^B+{\d_A}^B\pi_\ad)\ &=\ 0,\\
 \lambda^\ad(
  \nabla_{A\ad}\pi_\bd+(-)^{p_B}(R_{AB\ad\bd}-\Lambda_{AB}\epsilon_{\ad\bd})\omega^B)\ &=\ 0,
\end{aligned}
$$
i.e. they are annihilated by the local supertwistor connection
\eqref{eq:SC1}. Since the transformation $m_t\,:\,\lambda_\ad\mapsto t\lambda_\ad$
induces
$$ (\omega^A,\pi_\ad)\ \mapsto\ (t\omega^A,t\pi_\ad),$$
we conclude that the Penrose-Ward transform takes $\Scr{T}$ to
$\Scr{L}\otimes(T\Scr{L}/\{m_{t*}\})$, that is, to
$({\rm Jet}^1\Scr{L}^{-1})^\vee$.

\hfill $\Box$

This leads us to the following interesting result:

{\Pro\label{cor:pronatiso} 
There are natural isomorphisms of Berezinian sheaves:
\begin{itemize}
\setlength{\itemsep}{-1mm}
\item[(i)] ${\rm Ber}\Scr{J}\cong\Scr{L}^{-1}$,
\item[(ii)] ${\rm Ber}(\CP):={\rm Ber}\,\Omega^1\CP\cong\cL^{4-\CN}$.
\end{itemize}
}\vspace*{10pt} 

\noindent
{\it Proof:}  Starting point is the jet sequence \eqref{eq:jetseq}. This sequence 
in particular implies that
$$
 0\ \longrightarrow\ 
  \Omega^1\CP\otimes \Scr{L}^{-1}\ \longrightarrow\ {\rm Jet}^1\Scr{L}^{-1}\ \longrightarrow\ 
 \Scr{L}^{-1}\ \longrightarrow\ 0,
$$
i.e.
$$
 0\ \longrightarrow\ 
 \Scr{J}^\vee\ \longrightarrow\ {\rm Jet}^1\Scr{L}^{-1}\ \longrightarrow\ \Scr{L}^{-1}\ \longrightarrow\ 0,
$$
by virtue of Prop.\ \ref{pro:natiso}
Therefore, we obtain a natural isomorphism of
Berezinian sheaves\footnote{Note that ${\rm Ber}\,\Scr{L}\cong\Scr{L}$.}
$$ {\rm Ber}\Scr{J}\otimes\Scr{L}\ \cong\ {\rm Ber}\,({\rm Jet}^1\Scr{L}^{-1})^\vee. $$
Since\footnote{Recall that 
$\pi_2^*({\rm Jet}^1\Scr{L}^{-1})^\vee$ is free when restricted to $\pi_1^{-1}(x)$ for all
$x\in\CM$.} $$\pi_2^*(({\rm Jet}^1\Scr{L}^{-1})^\vee)\ \cong\ \pi_1^*\Scr{T}$$ 
and due to Eq.\ \eqref{eq:berLSTB}, we may conclude that 
$$  {\rm Ber}\,({\rm Jet}^1\Scr{L}^{-1})^\vee\ \cong\ \CO_\CP,$$
which, in fact, proves point (i). 

To verify point (ii), we merely apply Prop.\ \ref{pro:natiso} again. Indeed, from
$T\CP\cong\Scr{J}\otimes\Scr{L}^{-1}$ we find that
$$ {\rm Ber}(\CP)\ =\ {\rm Ber}\Scr{J}^\vee\otimes\Scr{L}^{3-\CN}\ =\ 
({\rm Ber}\Scr{J}\otimes\Scr{L})^\vee\otimes\Scr{L}^{4-\CN}\ \cong\ \Scr{L}^{4-\CN}.$$
This completes the proof.

\hfill $\Box$ 

\subsection{\kern-10pt Supertwistor space $(\CN=4)$}

Let us now discuss the $\CN=4$ case. However, we can be rather brief on this,
as the discussion is very similar to the one given above. Furthermore, for
the sake of illustration we only discuss the hyper-K\"ahler case. 

\paragraph{Conic structure and $\beta$-plane bundle.}
Let $\CM$ be a $(4|8)$-dimensional RC supermanifold equipped with the
Levi-Civita connection.
Recall again the sequence \eqref{eq:ES1}. It is equivalent to
\begin{equation}\label{eq:TBSN4}
 0\ \longrightarrow\ \CE[1]\otimes\widetilde\CS[-1]\ \longrightarrow\  
   T\CM\ \longrightarrow\ \CS[1]\otimes\widetilde\CS[-1]\ \longrightarrow\ 0.
\end{equation}
The reason for making this particular choice will become clear momentarily.

Let now $\CF$ be the relative projective line bundle 
$P_\CM(\widetilde\CS^\vee[1])$ on $\CM$. As before, the tangent bundle sequence 
induces a canonical $(2|4)$-conic structure on $\CM$, which in local coordinates
is given by
\begin{equation}\label{eq:canconN4}
\begin{aligned}
 \CF\ &\to\ G_\CM(2|4; T\CM),\\
 [\lambda_\ad]\ &\mapsto\ \Scr{D}\ :=\ \langle \lambda^\ad E_{A\ad}\rangle.
\end{aligned}
\end{equation}
By a similar reasoning as given in Prop.\ \ref{pro:PRF}, this distribution
will be integrable if and only if $\CM$ is right-flat. As before,
$\CF$ will be called the $\beta$-plane bundle in this case.
In addition,
Eq.\ \eqref{eq:lifteqre} is then substituted by
\begin{equation}\label{eq:lifteqreN4}
 \lambda^\ad D_{A\ad}\lambda_\bd\ =\ 0,
\end{equation}
i.e. $\lambda_\ad$ is auto-parallel
 with respect to the Levi-Civita connection on the $\beta$-plane
$\Sigma\hookrightarrow\CM$.
Furthermore, as directly follows from
the transformation laws given in Prop.\ \ref{pro:CoSLC},
this equation {\it is} scale invariant
since $\lambda_\ad$ is chosen to be a section of $\widetilde\CS^\vee[1]$. 
This explains,
why we have twisted $\widetilde\CS$ by $\CO_\CM[k]$, with $k=-1$
in \eqref{eq:TBSN4}.

\paragraph{Supertwistor space.}
As before, we obtain the following double fibration:
\begin{equation}\label{eq:DF3}
\begin{aligned}
\begin{picture}(50,40)
\put(0.0,0.0){\makebox(0,0)[c]{$\CP$}}
\put(64.0,0.0){\makebox(0,0)[c]{$\CM$}}
\put(34.0,33.0){\makebox(0,0)[c]{$\CF$}}
\put(7.0,18.0){\makebox(0,0)[c]{$\pi_2$}}
\put(55.0,18.0){\makebox(0,0)[c]{$\pi_1$}}
\put(25.0,25.0){\vector(-1,-1){18}}
\put(37.0,25.0){\vector(1,-1){18}}
\end{picture}
\end{aligned}
\end{equation}
Here, $\CP$ is the $(3|4)$-dimensional supertwistor
space of $\CM$. Again, we need to assume that $\CM$ is civilized.

As already indicated, we shall now directly
jump to the hyper-K\"ahler case. The following then gives
the inverse construction.
{\Thm\label{thm:NLGN4} 
There is a one-to-one correspondence between civilized
complex hyper-K\"ahler supermanifolds $\CM$ of dimension
$(4|8)$ and
$(3|4)$-dimensional complex supermanifolds $\CP$ such that:
\begin{itemize}
\setlength{\itemsep}{-1mm}
\item[(i)] $\CP$ is a holomorphic fibre bundle $\pi\,:\,\CP\to\IP^1$
           over $\IP^1$,
\item[(ii)] $\CP$ is equipped with a $(4|8)$-parameter family of
            sections of $\pi$, each with normal bundle $\cN_{\IP^1|\CP}$ described by 
             $$
            0\ \longrightarrow\ \Pi\CO_{\IP^1}(1)\otimes\IC^4\ \longrightarrow\  
                  \cN_{\IP^1|\CP}\ \longrightarrow\ \CO_{\IP^1}(1)\otimes\IC^2\ \longrightarrow\ 0,
             $$
            and
\item[(iii)] there exists a holomorphic relative symplectic structure $\omega$ of
             weight $2$ on $\CP$.
\end{itemize}
}

In proving this result, one basically follows the argumentation given in
Sec.\ \ref{sec:STC} The only modification is the replacement of
$\widetilde\CS$ by  $\widetilde\CS[-1]=\widetilde\CS\otimes{\rm Ber}\,\widetilde\CS$. 
In this respect, we also point out that triviality of the bundle
$\widetilde\CS[-1]$ 
certainly implies triviality of $\widetilde\CS$.

\paragraph{Remark.}
It is obvious, how to define the universal line bundle, the Jacobi bundle and
the bundle of local supertwistors in 
the context of the $\CN=4$ supertwistor space. Prop.\ \ref{pro:natiso} can be modified
accordingly. Point (ii) of Prop.\ \ref{cor:pronatiso} is then substituted by the fact that the
Berezinian sheaf Ber$(\CP)$ is globally trivial, i.e.  Ber$(\CP)\cong\CO_\CP$.
Hence, the $\CN=4$ supertwistor space is a formal Calabi-Yau supermanifold.

\section{\kern-10pt Real structures}\label{sec:RS}

So far, we have been discussing only complex supermanifolds. The 
subject of this section
is to comment on a real version of theory. 

\paragraph{Almost quaternionic supermanifolds.}\label{par:AQS}
Let us first present an overview about {\it real structures} on 
complex supermanifolds.

{\Def [Manin \cite{Manin}]
A real structure on a complex supermanifold $(\CM,\CO_\CM)$ of
type $(\epsilon_1,\epsilon_2,\epsilon_3)$, where $\epsilon_i=\pm1$
for $i=1,2,3$, is an even $\IR$-linear mapping $\rho\,:\,\CO_\CM\to\CO_\CM$
such that
$$ \rho(\a f)\ =\ \bar{\alpha}\rho(f),\qquad
   \rho(\rho(f))\ =\ (\epsilon_1)^{p_f}f,\qquad
   \rho(fg)\ =\ \epsilon_3(\epsilon_2)^{p_fp_g}\rho(g)\rho(f), $$
where $f$ and $g$ are local holomorphic functions on $\CM$ and $\alpha\in\IC$.
The bar means complex
conjugation. Furthermore, $\rho(f(\cdot))=\overline{f(\rho(\cdot))}$.

If $\CE$ is a holomorphic vector bundle over $\CM$, then a prolongation
$\hat\rho$ of type $\eta=\pm1$ of a given real structure $\rho\,:\,\CO_\CM\to\CO_\CM$
is an even $\IR$-linear mapping $\hat\rho\,:\,\CE\to\CE$ such that
$$\hat\rho(\hat\rho(\sigma))\ =\ \eta(\epsilon_1)^{p_\sigma}\sigma,\qquad
  \hat\rho(f\sigma)\ =\ \epsilon_3(\epsilon_2)^{p_fp_\sigma}\hat\rho(\sigma)\rho(f),\qquad
  \hat\rho(\sigma f)\ =\ \epsilon_3(\epsilon_2)^{p_fp_\sigma}\rho(f)\hat\rho(\sigma),$$
where $\sigma$ is a local section of $\CE$ and $f$ is a local holomorphic function
on $\CM$. If $\eta=+1$ then the prolongation is called real while for $\eta=-1$ 
quaternionic.}\vspace*{10pt}

Having recalled the definition of real structures and their extensions to
vector bundles, we may now give the following definition:

{\Def\label{def:QRS}
A $(4|2\CN)$-dimensional RC supermanifold $\CM$ is called an almost 
quaternionic RC supermanifold 
if there is a real structure $\rho$ on $\CM$ of
type $(-1,1,1)$ which leaves $\CE\otimes\widetilde\CS$ invariant and 
which induces
two quaternionic prolongations $\hat\rho_1\,:\,\CS\to\CS$ and
$\hat\rho_2\,:\,\widetilde\CS\to\widetilde\CS$, respectively. 
In addition, it is also assumed
that $\rho$ has a (real) $(4|2\CN)$-dimensional supermanifold 
$\CM_\rho$ of $\rho$-stable points in $\CM$.}\vspace*{10pt}  

\paragraph{Structure group on $\CM_\rho$.}
In \ref{par:SG}~~, we have discussed the form of the structure group $G$ of $T\CM$. If $\CM$
is equipped with an almost quaternionic structure, $G$ may be reduced  to the real form
$G_\rho$ on $\CM_\rho$ which is described by\footnote{See Salamon \cite{Salamon:1982} 
for the purely bosonic situation.}
\begin{equation}\label{eq:coverreal}
 1\ \longrightarrow\ \IZ_{|4-\CN|}\ \longrightarrow\ 
 S(GL(1|\tfrac{1}{2}\CN,\IH)\times GL(1|0,\IH))\ \longrightarrow\ G_\rho\ \longrightarrow\ 1,
\end{equation}
where $G_\rho\subset GL(4|2\CN,\IR)$. This makes it clear why it is necessary
to have an {\it even} number $\CN$ of supersymmetries 
as otherwise one cannot endow an RC supermanifold with an almost quaternionic structure. 

Furthermore, a scale is defined in this case  as
follows:
{\Def A scale on an almost quaternionic RC supermanifold $\CM$ is a choice of a particular
non-vanishing volume form $\tilde\varepsilon\in H^0(\CM,{\rm Ber}\,\widetilde\CS^\vee)$
on the vector bundle $\widetilde\CS$ such that the corresponding volume form
${\rm Vol}\in H^0(\CM,{\rm Ber}(\CM))$ obeys
$\rho({\rm Vol})= {\rm Vol}$.
}\vspace*{10pt}

\noindent
Clearly, a choice of scale reduces the structure group $G_\rho$ on $\CM_\rho$ further down to 
$SG_\rho$, which in fact is given by
\begin{equation}\label{eq:coverreal3}
 1\ \longrightarrow\ \IZ_2\ \longrightarrow\ 
 SU(2|\CN)\times SU(2|0)\ \longrightarrow\ SG_\rho\ \longrightarrow\ 1,
\end{equation}
as follows from \eqref{eq:cover1}.

Now one can basically repeat the analysis given in Secs.\ \ref{sec:pre} and \ref{sec:sdsg}
starting from almost quaternionic RC supermanifolds. One eventually arrives
at the notions of quaternionic, quaternionic K\"ahler and hyper-K\"ahler structures,
that is, in Defs.\ \ref{def:CQRC}, \ref{def:CQKRC} and \ref{def:CHKRC} one simply
needs to remove the word ``complex".

\paragraph{Supertwistor space.}
It remains to clarify the additional structure on the supertwistor space
$\CP$ needed in order to be associated with an RC supermanifold equipped with
a real structure in the above sense. 

On first notices that by starting from $\CM$, the real structure
$\rho$ on $\CM$ naturally induces real structures on $\CF$ and $\CP$, respectively, which are, of
course, of the same type as $\rho$, that is, $(-1,1,1)$. For instance, since $\rho$ is assumed to have 
a quaternionic prolongation $\hat\rho_2\,:\,\widetilde\CS\to\widetilde\CS$, the induced real
structure acts on the
fibres of $\pi_1:\CF\to\CM$ as the antipodal map 
$(\lambda_{\dot1},\lambda_{\dot2})\mapsto(-\bar\lambda_{\dot2},\bar\lambda_{\dot1})$.
Since $\CP$ foliates $\CF$, one obtains the induced real structure on $\CP$.
The following theorem clarifies also the reverse direction:

{\Thm\label{thm:T2}
There is a one-to-one correspondence between:
\begin{itemize}
\setlength{\itemsep}{-1mm}
\item[(i)] civilized  right-flat quaternionic RC supermanifolds $\CM$ of (complex)
           dimension $(4|2\CN)$ and
\item[(ii)] $(3|\CN)$-dimensional complex supermanifolds $\CP$ each containing a 
            family of 
            holomor\-phically embedded projective lines $\IP^1$ each having normal
            bundle $\cN_{\IP^1|\CP}$ inside $\CP$  described by 
            \eqref{eq:normalbundle}
            and in addition, $\CP$ has a real structure of type
            $(-1,1,1)$ which is compatible with the above data and which acts on the projective lines 
            $\IP^1$ as the antipodal map.
\end{itemize}
}\vspace*{10pt}

\noindent
{\it Proof:} 
In fact, almost everything has been proven (cf. also Thm.\ \ref{thm:T1}).
It remains to show that by going from (ii) $\to$ (i) the antipodal map on $\CP$
indeed gives the correct real structure on $\CM$.
To see that the induced real
structure $\rho$ on $\CM$ yields two quaternionic prolongations $\hat\rho_1\,:\,\CS\to\CS$
and  $\hat\rho_2\,:\,\widetilde\CS\to\widetilde\CS$, respectively, we 
apply arguments of Hitchin et al. \cite{Hitchin:1986ea}. 

In particular, consider $\widetilde\CS=\pi_{1*}(\pi_2^*\CO_{\IP^1}(1))$.  
Then the prolongation $\hat\rho_2$, induced by the antipodal map on $\IP^1$,
is given by
$$ \hat\rho_2(a_\ad\lambda^\ad)\ :=\ \bar a_{\dot2}\lambda^{\dot1}-\bar a_{\dot 1}\lambda^{\dot2}.$$
Analogously, the antipodal map induces a quaternionic prolongation $\hat\rho_1$ on 
the bundle $\CS=\pi_{1*}(\pi_2^*(\CO_{\IP^1}\otimes\IC^2))$.

\hfill $\Box$ 

\noindent
In a similar fashion, one may make the appropriate changes in 
Thm.\ \ref{thm:NLG}

Finally, we have the following fact:

{\Pro\label{pro:diffeo} Let $\CM$ be a civilized right-flat
quaternionic RC supermanifold and $\CP$ its associated 
supertwistor space. Then 
there is a natural diffeomorphism $\CP\cong P_{\CM_\rho}(\widetilde\CS^\vee|_{\CM_\rho})$. Hence,
we obtain a nonholomorphic fibration
$$ \CP\ \to\ \CM_\rho$$
of the supertwistor space over $\CM_\rho\subset\CM$. Typical fibres of this fibration
are two-spheres $S^2$.} 
\vspace*{10pt}

\paragraph{Remark.}
In the purely bosonic setting and for Euclidean signature, 
the twistor space has an alternative definition which is
equivalent to the definition in terms of the projectivization of the right-chiral
spin bundle.
Let $M$ be an oriented Riemannian four-manifold. The twistor space $P$ of
$M$ can equivalently be defined as the associated bundle (Atiyah et al. \cite{Atiyah:1978wi})
\begin{equation}
 P\ :=\ P(M,SO(4))\times_{SO(4)}(SO(4)/U(2))
\end{equation}
with
\begin{equation}\label{eq:2.2}
 P\ \to\ M.
\end{equation}
Typical fibres of this bundle are two-spheres $S^2\cong SO(4)/U(2)$ which parametrize
almost complex structures on the fibre $T_xM$ of $TM$ over $x\in M$.
Recall that an almost complex structure $\mathcal{J}$ is an endomorphism
of the tangent bundle that squares to minus the identity, i.e.
$\mathcal{J}^2=-1$.
Note that while a manifold $M$ admits in general no almost complex
structure, its twistor space $P$ can always be equipped with an
almost complex structure $\mathcal{J}$ (Atiyah et al. \cite{Atiyah:1978wi}). Furthermore,
$\mathcal{J}$ is integrable if and only if the Weyl tensor of $M$ is
self-dual \cite{Penrose:in,Atiyah:1978wi}. Then $P$ is a complex
three-manifold with an antiholomorphic involution $\rho$ 
which maps $\mathcal{J}$ to $-\mathcal{J}$ and the fibres of the bundle
\eqref{eq:2.2} over $x\in M$ are $\rho$-invariant projective lines
$\IP^1$, each of which has normal bundle $\CO_{\IP^1}(1)\otimes\IC^2$
in the complex manifold $P$. Here and in the
following we make no notational distinction between real structures 
appearing on different (super)manifolds.

In the supersymmetric setting, the situation is slightly different. Let
us consider $\CM_\rho$ from above. The tangent spaces $T_x\CM_\rho$ 
for $x\in\CM_\rho$ are isomorphic to $\IR^{4|2\CN}$. So almost complex
structures are parametrized by the supercoset space\footnote{For more details, 
see e.g. Wolf \cite{Wolf:2004hp}.} $OSp(4|2\CN)/U(2|\CN)$,
which is a supermanifold of (real) dimension $2+\CN(\CN+1)|4\CN$, and whose
even part is\footnote{Recall that if $G$ is a Lie supergroup and
$H$ a closed Lie subsupergroup (i.e. $H_{\rm red}$ is closed in $G_{\rm red}$)
then $G/H:=(G_{\rm red}/H_{\rm red},\CO_{G/H})$, where $\CO_{G/H}(\CU):=
\{f\in\CO_G(\pi^{-1}(\CU))\,|\,\tilde\phi f={\rm pr}^* f\}$ with
$\CU\subset H_{\rm red}$, $\pi\,:\,G_{\rm red}\to G_{\rm red}/H_{\rm red}$
and pr$\,:\,G\times H\to G$ are the canonical projections and 
$\varphi=(\phi,\tilde\phi)\,:\,G\times H\to G$ is the right
action of $H$ on $G$. See Kostant \cite{Kostant:1977} for more details. Hence,
$(G/H)_{\rm red}\equiv G_{\rm red}/H_{\rm red}$.}
\begin{equation}
 (SO(4)\times Sp(2\CN,\IR))/(U(2)\times U(\CN))\ \cong\ 
  SO(4)/U(2)\times Sp(2\CN,\IR)/U(\CN).
\end{equation}
Thus, the supertwistor space $\CP\to\CM_\rho$, as viewed as in
Prop.\ \ref{pro:diffeo}, cannot be 
reinterpreted as a 
space which does describe all possible almost complex structures on $\CM_\rho$.
Nevertheless, one can view $\CP$ as a space describing a certain
class of almost complex structures on $\CM_\rho$. Remember that the complexified
tangent bundle $T\CM_\rho\otimes\IC$ can be factorized as
$T\CM_\rho\otimes\IC\cong\CH\otimes\widetilde\CS$. In particular,
these complex structures, being compatible with this tangent bundle
structure, are again parametrized by two-spheres
$S^2\cong SO(4)/U(2)$ and are given in a structure frame by (cf. Wolf
\cite{Wolf:2004hp})
\begin{equation}\label{eq:cplstr}
 {\mathbf{J}_{A\ad}}^{B\bd}\ =\ -{\rm i}{\d_A}^B\frac{\lambda_\ad\hat\lambda^\bd+
 \lambda^\bd\hat\lambda_\ad}{\lambda_\gd\hat\lambda^\gd},
\end{equation}
where $\lambda_\ad$ are homogeneous coordinates on $\IP^1$ ($\cong S^2$) and $^t(\hat\lambda^\ad):=(\bar\lambda^{\dot2},-\bar\lambda^{\dot1})$ (see also
the preceding paragraph). Now one may introduce an almost complex structure
$\mathcal{J}$ on $\CP\overset{S^2}{\to}\CM_\rho$ by setting $\mathcal{J}_z=\mathbf{J}_z\oplus J_z$
for $z\in\CP$. Here, $\mathbf{J}_z$ is given in terms of \eqref{eq:cplstr}
and $J_z$ in terms of the standard almost complex structure on $S^2$, respectively.
In fact, following the arguments of Atiyah et al. \cite{Atiyah:1978wi},
this description of $\mathcal{J}$ does not depend on the choice of local
coordinates. Hence, $\mathcal{J}_z$ can be defined for all $z\in\CP$ and
thus, $\CP$ comes equipped with a natural almost complex structure.

Next one can show that this almost complex
structure is integrable if and only if $\CM$ is right-flat and furthermore
that the fibres of $\CP\to\CM_\rho$ are $\rho$-invariant projective lines $\IP^1$ each
having normal bundle $\cN_{\IP^1|\CP}$ inside $\CP$ described by 
\begin{equation}
            0\ \longrightarrow\ \Pi\CO_{\IP^1}(1)\otimes\IC^\CN\ \longrightarrow\  
                  \cN_{\IP^1|\CP}\ \longrightarrow\ \CO_{\IP^1}(1)\otimes\IC^2\ \longrightarrow\ 0.
\end{equation}

\vspace*{.5cm}

\begin{center}{\large\sc Acknowledgments}\end{center}

\vspace*{.2cm}

\noindent
I would like to thank Paul Howe, Chris Hull, Olaf Lechtenfeld, Lionel Mason, Alexander Popov, 
Riccardo Ricci, Christian S\"amann and Robert Wimmer
for a number of very useful discussions. In particular, I am very grateful to
Alexander Popov for drawing my attention to this project, for
sharing his ideas and for collaborating at an early stage. This work
 was supported in part by the EU under the MRTN contract MRTN--CT--2004--005104
and by PPARC under the rolling grant PP/D0744X/1.

\vspace*{.5cm}

\begin{center}{\large\sc Appendix}\end{center}

\renewcommand{\thesection}{\Alph{section}.}
\setcounter{subsection}{0} \setcounter{equation}{0}
\renewcommand{\thesubsection}{\Alph{subsection}}
\renewcommand{\theequation}{\thesubsection.\arabic{equation}}

\subsection{\kern-10pt Proof of Prop.\ \ref{pro:curv}}\label{app:curv}

Subject of this appendix is to give a proof of Prop.\ \ref{pro:curv}
First, we show the second relation of Eqs.\ \eqref{eq:procurv}. The proof
of the third one follows similar lines as for the second one. So we
omit it at this point
and leave it to the reader. Eventually, we prove the first relation. 

The curvature components ${R_{ABC}}^D$ can be decomposed into irreducible\footnote{Note that
in order to obtain the set of independent superfield components, one has to go 
one step further and employ the second Bianchi identity
(see e.g. Eq.\ \eqref{eq:BI}).}
pieces as
\begin{equation}\label{eq:curvirred}
\begin{aligned}
 {R_{ABC}}^D\ =\ {C_{ABC}}^D &+ {D_{ABC}}^D+E_{AB}{\d_C}^D\ +\\
    &+\ (\CN-2)(-)^{p_C(p_A+p_B)}
 E_{C\{A}{\d_{B]}}^D-2(-)^{p_C(p_A+p_B)}\Lambda_{C\{A}{\d_{B]}}^D,
\end{aligned}
\end{equation}
where ${C_{ABC}}^D$ and $\Lambda_{AB}$ obey the properties stated in
Prop.\ \ref{pro:curv} and ${D_{ABC}}^D={D_{\{AB]C}}^D$, ${D_{\{ABC]}}^D=0$
and $E_{AB}=E_{\{AB]}$. Furthermore, ${D_{ABC}}^D$ is totally trace-free.

Recall the Bianchi identity 
$${R_{[A\ad B\bd C\gd\}}}^{D\dd}\ =\ 0,$$ which reads 
explicitly as
\begin{equation}
 {R_{A\ad B\bd C\gd}}^{D\dd}+(-)^{p_A(p_B+p_C)}{R_{B\bd C\gd A\ad}}^{D\dd}
 +(-)^{p_C(p_A+p_B)}{R_{C\gd A\ad B\bd}}^{D\dd}\ =\ 0.
\end{equation}
Upon substituting
\begin{equation}
 {R_{A\ad B\bd C\gd}}^{D\dd}\ =\ \big[\epsilon_{\ad\bd}{R_{ABC}}^D+{R_{A(\ad B\bd)C}}^D\big]{\d_\gd}^\dd
                                +\big[\epsilon_{\ad\bd}{R_{AB\gd}}^\dd+{R_{A(\ad B\bd)\gd}}^\dd\big]{\d_C}^D,
\end{equation}
which follows from \eqref{eq:curvdec} and upon contracting with $\epsilon_{\dd\dot\epsilon}$, we
arrive at
\begin{equation}\label{eq:permcurv}
\begin{aligned}
 &\big[\epsilon_{\ad\bd}{R_{ABC}}^D+{R_{A(\ad B\bd)C}}^D\big]\epsilon_{\gd\dd}+
 \big[\epsilon_{\ad\bd}{R_{AB\gd\dd}}+{R_{A(\ad B\bd)\gd\dd}}\big]{\d_C}^D\ +\\
&\kern.5cm+\ (-)^{p_A(p_B+p_C)}\big[\epsilon_{\bd\gd}{R_{BCA}}^D+{R_{B(\bd C\gd)A}}^D\big]\epsilon_{\ad\dd}+
 \big[\epsilon_{\bd\gd}{R_{BC\ad\dd}}+{R_{B(\bd C\gd)\ad\dd}}\big]{\d_A}^D\ +\\
&\kern1cm+\ (-)^{p_C(p_A+p_B)}\big[\epsilon_{\gd\ad}{R_{CAB}}^D+{R_{C(\gd A\ad)B}}^D\big]\epsilon_{\bd\dd}+
 \big[\epsilon_{\gd\ad}{R_{CA\bd\dd}}+{R_{C(\gd A\ad)\bd\dd}}\big]{\d_B}^D\ =\ 0.\\
\end{aligned}
\end{equation}

Therefore, upon looking at the terms proportional to $\epsilon_{\ad\bd}\epsilon_{\gd\dd}$
(plus a permutation of the indices),
one arrives after some lengthy but straightforward calculations at
\begin{equation}\label{eq:traceprop}
 (-)^C {R_{ABC}}^C\ =\ 0\qquad{\rm and}\qquad (-)^C {R_{\{ABC]}}^C\ =\ 0.
\end{equation}
In addition,
\begin{equation}
\begin{aligned}
  {R_{ABC}}^D\ &=\ \tfrac{1}{3}({R_{ABC}}^D+(-)^{p_Bp_C}{R_{ACB}}^D+(-)^{p_A(p_B+p_C)}{R_{BCA}}^D)\ +\\
                  &\kern1cm+\ \tfrac{1}{3}({R_{ABC}}^D-(-)^{p_Bp_C}{R_{ACB}}^D)\ +\\
                  &\kern2cm+\ \tfrac{1}{3}({R_{ABC}}^D-(-)^{p_A(p_B+p_C)}{R_{BCA}}^D)\\
            &=\ {R_{\{ABC]}}^D+\tfrac{2}{3}{R_{A[BC\}}}^D+\tfrac{2}{3}{R_{B[AC\}}}^D.
\end{aligned}
\end{equation}
By comparing this result with Eqs.\ \eqref{eq:curvirred} and \eqref{eq:traceprop}, we conclude
that ${D_{ABC}}^D$ and $E_{AB}$ must vanish and ${R_{\{ABC]}}^D={C_{ABC}}^D$. Hence,
 \begin{equation}
 {R_{ABC}}^D\ =\ {C_{ABC}}^D -2(-)^{p_C(p_A+p_B)}\Lambda_{C\{A}{\d_{B]}}^D,
\end{equation}
which is the desired result.

Let us now discuss the first relation given in Eqs.\ \eqref{eq:procurv}.
By looking at terms in Eq.\ \eqref{eq:permcurv} which are symmetric in 
$\ad,\bd$ but antisymmetric in $\gd,\dd$, we find that
\begin{equation}
\begin{aligned}
 &\big[{R_{A(\ad B\bd)C}}^D+2(-)^{p_C(p_A+p_B)}R_{C[A\ad\bd}{\d_{B\}}}^D\big]\ +\\
 &\kern.5cm+\ \big[{R_{A(\ad B\bd) C}}^{D}+(-)^{p_A(p_B+p_C)}{R_{B(\ad C\bd) A}}^{D}
 +(-)^{p_C(p_A+p_B)}{R_{C(\ad A\bd) B}}^{D}\big]\ =\ 0.
\end{aligned}
\end{equation}
However, the second line vanishes identically as it represents a  
Bianchi identity for the curvature of the bundle $\CH\to\CM$.
Therefore, we end up with
\begin{equation}
 {R_{A(\ad B\bd)C}}^D\ =\ -2(-)^{p_C(p_A+p_B)}R_{C[A\ad\bd}{\d_{B\}}}^D.
\end{equation}

Finally, we notice that the form \eqref{eq:riccitensor} of the Ricci tensor 
can straightforwardly be
obtained by substituting Eqs.\ \eqref{eq:procurv} into its definition
and by explicitly performing the 
appropriate index traces. This remark concludes the proof of
Prop.\ \ref{pro:curv}

\end{document}